\documentclass{article}

\usepackage{fullpage}
\usepackage{amsmath}
\usepackage{textcomp}
\usepackage{amsthm}
\usepackage{listings}
\usepackage{xcolor}
\usepackage{graphicx}
\usepackage{float}
\usepackage{cancel}

\lstdefinestyle{cppstyle}{
    language=C++,
    backgroundcolor=\color{lightgray!20},
    basicstyle=\ttfamily\footnotesize,
    keywordstyle=\color{blue},
    stringstyle=\color{orange!60!black},
    commentstyle=\color{teal},
    numberstyle=\tiny\color{black},
    numbers=left,
    stepnumber=1,
    frame=single,
    breaklines=true,
    tabsize=4,
    showspaces=false,
    showstringspaces=false,
}

\newtheorem{theorem}{Theorem}
\newtheorem{lemma}{Lemma}
\newtheorem{corollary}{Corollary}
\newtheorem{observation}{Observation}

\title{Algorithmic Structure In Subset Sum: Deterministic In-Bound Navigation And The Counting Complexity Divide}
\author{Thami Nkosi}

\begin{document}
\maketitle

\begin{abstract}
    This paper presents a deterministic algorithmic approach of exploring the solution space of the Subset Sum Problem. The algorithm presented is input-robust and structurally adaptive. Exploration is guided and narrows into areas in the solution space where solutions are possible, referred to as in-bound solution space, skipping all areas where solutions are impossible. Unfortunately, this can lead to false positives: paths that are hinted to potential have solutions but ultimately realized to not lead to solutions. The in-bound solution space navigated can therefore be filled with only false positives, only true solutions or a mix of the two, affecting the algorithm's performance in different ways. We then detail the challenges of exploring the in-bound solution space for different instances. Further, we show how this algorithm may practically generalize to other NP/NP-complete problems with appropriate adaptation. An introductory discussion is done on this generalization to k-SAT and general CNF-SAT, deferring extensive detail to a follow-up paper. This paper does not satisfy P vs NP proof requirements and does not claim to resolve the problem. However, it has implications for the P vs NP and offers a practical lens through the algorithm of what is feasible with it. The feasibility bounds of the algorithm reveal a nontrivial relationship between decision and counting complexity. To facilitate easy reproducibility, we include in the paper a full C++ implementation of the algorithm.
\end{abstract}



\pagebreak

\section{Introduction}
The Subset Sum Problem is a well-known problem in combinatorial optimization and computational complexity theory. As it is an NP-complete problem, it has the nature of being computationally challenging to solve as the input size increases. Specifically, the number of possible subsets of the problem increase exponentially with number of elements in a given set. Any problem in the NP class theoretically can be reduced to a Subset Sum Problem in polynomial time as it is NP-complete and harder than or equal in complexity to all problems in NP\cite{cook1971}.

The problem has ancient roots in mathematical challenges and serves as a classic NP-complete problem. Any subset proposed as a solution to an instance of the Subset Sum Problem can be verified in polynomial time, but a subset that is a solution cannot be deterministically found in polynomial time for all instances. In a typical Subset Sum Problem, you are given a set of integers, can be positive, negative or 0, and a target sum. The task is to then determine if there exists a subset/s of the given set that sums to the target sum. Variants of the problem are the decision version, where the aim to only determine if a subset that sums to the target sum exists or not, and the counting version, where the aim is to count the number of subsets that sum to the target sum. Other variations also exist with different constraints.

\begin{itemize}
    \item \textbf{Definition of Key Terms In The Paper:}
    \begin{enumerate}
        \item \textbf{Binary Increment Enumeration}: Can also be referred to as "binary increments of a set" in the paper. This describes the systematical way of generating subsets of a set by binary encodings of an (n)-bit binary string. Enumeration follows the binary increments of the (n)-bit binary string from 000\ldots0 to 111\ldots1. Each bit corresponds to each of the '$n$' elements in the set.
        \item \textbf{Possibility Space}: This refers to all combinations/subsets that can be made from a given set. Can also be referred to as solution space
        \item \textbf{Possibility}: Refers to a single combination or subset that can be achieved by a given set. In the paper, the words subset, combination, and possibility will be used interchangeably, but they refer to the same thing.
        \item \textbf{S/E Point(s)}: In full, starting and ending points. These are only specifically chosen possibilities in the possibility space that have varying range of possibilities between them. These help us navigate the possibility space efficiently to find the needed target sum.
        \item \textbf{Zoom-In}: This is the process of further determining S/E points within other S/E points to narrow down exactly where the possibilities that equal to the target sum might be in the possibility space.
        \item \textbf{In-Bound Solution Space}: This is the collective name of all areas of the possibility space that we zoom into guided by the target sum criteria hinting solutions. See \textbf{Candidate(s)} below.
        \item \textbf{False Positive(s)}: In-bound solution space paths we traverse that are initially hinted to have solutions and are ultimately realized to not lead to solutions. We consider an exploration a false positive only if the algorithm traverses the full path till the end before it discards it of not leading to solutions. Otherwise, the algorithm would discard the path before reaching the end of it. These are only wasteful path traversals not leading to solutions and not necessarily incorrect solution given back by the algorithm.
        \item \textbf{Candidate(s)}: Refers to the outcome we get at the end of paths we traverse. The outcome at the end of every path can either be a false positive or true solution. A possibility is only counted/considered a candidate if the target sum criteria leads us to it at the end of a path.\\
        \textbf{In-bound Solution Space}, \textbf{False Positives}, and \textbf{Candidates}, are synonymous. In-bound solution space is made up by candidates, and false positives a kind of candidates.
        \item \textbf{Global Fluctuation}: In a plot of subsets of a set, specifically in mixed-sign sets. The plot of subsets would be in a 2D of the negative and positive sign. This would refer to the fluctuation of the outer sign set in the 2D plot of subsets.
        \item \textbf{Local Fluctuation}: Same thing as global fluctuations in the 2D plot of the subsets, in this case, this refers to the fluctuation of the inner sign set.
        \item \textbf{Z-Axis Exponentiation}: Describes the exponential increase in number of layers of identical xy-plane plot of subsets placed on top of each other caused by the presence of zeros in a given set.
        \item \textbf{Literal Mapping}: This is the conversion of variables/literals and their negations in a SAT formula to decimal values.
        \item \textbf{Clause Value}: A value determined by adding up the literal values in a clause according to the literal mapping.
        \item \textbf{Root Clause}: A clause used to give us a first hint/lead to a satisying assignment solution of a SAT instance.
        \item \textbf{Clause Signature}: Identifies a clause by the specific literals and number of literals in it.
        \item \textbf{Internal Solutions}: Satisfying assignment solutions to a SAT instance that we get from a root clause that is part of the formula.
        \item \textbf{External Solutions}: Satisfying assignment solutions to a SAT instance that we get from a root clause that is not part of the formula.
    \end{enumerate}
    \item \textbf{Definition of Key Notation In The Paper:}
    \begin{center}
        Number of Preceding Possibilities (Starting, Ending)\textsubscript{Distance} / N (S, E)\textsubscript{D}
    \end{center}
    \begin{enumerate}
        \item \textbf{Starting/S}: A specifically chosen, known, and calculated possibility in the possibility space that serves as a start of a range of unknown possibilities that come after it.
        \item \textbf{Ending/E}: A specifically chosen, known, and calculated possibility in the possibility space that serves as an end of a range of unknown possibilities that come before it.
        \item \textbf{Distance/D}: The number of unknown possibilities that are between the known starting and ending point plus 1 (the ending point). The number of unknown possibilities includes the count of the known ending point but does not count the starting point.
        \item \textbf{Number of Preceding Possibilities/N}: Total number of possibilities that come before the starting point. We do not count the starting point itself. S/E points that are within other S/E points inherit this number from their respective parent/base S/E point and continue from it.
    \end{enumerate}
\end{itemize}

\textbf{Every subsection in the methodology section that has an asterisk (*) affects/contributes to the overall complexity of the method to be presented in the paper and the overall complexity of the method will be fully analyzed in subsection 3.13.}

The Subset Sum problem has significance in fields like cryptography, finance, resource allocation, Artificial Intelligence and Machine Learning, and other fields where finding a specific subset with certain properties is important. In what follows, a way of exploring the exponentially growing subsets efficiently is examined in detail. Also included in the paper, is the method implementation coded in C++.


\section{Related Work}
Methods such as dynamic programming\cite{bellman1952}, approximation algorithms\cite{johnson1974}, heuristics, and even a hybrid of those approaches are the current best available methods when it comes to solving NP-complete problems, including the Subset Sum Problem. All these methods have instances where they function efficiently, but have limitations for larger instances and other edge cases. For example, dynamic programming functions well with small to moderate size target sum instances of the Subset Sum Problem, but takes exponential time to solve instances with large target sums.

The method presented in this paper improves on the shortcomings of the known methods, its bounds only governed by the actual structural complexity of an instance. It has been tested also in edge cases including a mix of both positive and negative numbers in a given set, making it a great alternative over current methods. The algorithm also provides capabilities of mapping back solutions' elements, naturally parallelizable computation of paths traversed, incremental solution count and multiple target sum searching on-the-fly, in a single run.


\section{Methodology}
Now to unpack the algorithm, because it involves exploring the actual combinations or subsets, we first have to establish a way of enumerating the combinations in a way that would reveal exploitable sequences and patterns of the combinations. In simple, we enumerate the possibilities in binary increments of the given set. Given the set: \{1, 2, 3, 4\}, we first enumerate 0 as it is the empty set possibility. Next, we take the first element of the set and place it as a possibility in the enumeration after 0. Then we add the element to each of the preceding possibilities between 0 and the current element. Each sum to a preceding possibility is added/enumerated as a possibility itself after the element we are adding.

Of course, this would not work with only 0 and the first element of the set, because the first possibility after 0 is the element itself. So, we continue and take the next element from the set and add it as a possibility and again, add the element to the preceding possibilities except 0 and itself. We keep on moving to the next element of the set until we have added all the elements of the set as possibilities and performed the addition. Let us see this:

\begin{enumerate}
    \item \{1, 2, 3, 4\}: 0, \textbf{(1)}, \textbf{(2)}, 3, \textbf{(3)}, 4, 5, 6, \textbf{(4)}, 5, 6, 7, 7, 8, 9, 10.
    \item \{2, 4, 6, 8\}: 0, \textbf{(2)}, \textbf{(4)}, 6, \textbf{(6)}, 8, 10, 12, \textbf{(8)}, 10, 12, 14, 14, 16, 18, 20.
    \item \{1, 3, 5, 7, 9\}: 0, \textbf{(1)}, \textbf{(3)}, 4, \textbf{(5)}, 6, 8, 9, \textbf{(7)}, 8, 10, 11, 12, 13, 15, 16, \textbf{(9)}, 10, 12, 13, 14, 15, 17, 18, 16, 17, 19, 20, 21, 22, 24, 25.
    \item[] This can also be done for negative numbers or a mix of 0, negative and positive numbers:
    \item \{-4, -2, 0, 2\}: 0, \textbf{(-4)}, \textbf{(-2)}, -6, \textbf{(0)}, -4, -2, -6, \textbf{(2)}, -2, 0, -4, 2, -2, 0, -4.
\end{enumerate}

    
    \subsection{Binary Increments Of Any Set Enumerating All possible Combinations Of The Set}

    \begin{lemma}
        Binary increments of any given set, regardless of the nature or sign of the elements in the set, enumerates all possible combinations that can be achieved by the set.
        
        \begin{enumerate}
            \item The total number of subsets achievable by a set of elements is $2^n$, where '$n$' is the number of elements. Each element in the set can be either included or excluded in a subset. Two choices (inclusion or exclusion) can be made for each element in the set and therefore, 2 choices multiplied/made '$n$' number of times.
            \item Any subset of set $S$ can be represented as a binary sequence of length $n$, where $n$ is the number of elements in $S$. Each bit in the binary sequence corresponds to an element and shows us whether the corresponding element in $S$ is included or excluded in the represented subset. Two choices for each bit to be set to 1 or 0, and for all bits an overall of 2 choices must be multiplied/made '$n$' number of times.
            \item Binary increments enumerate all binary strings of length n, starting from 000\ldots0 to 111\ldots1. Each element in $S$ that corresponds to a “1” in the binary string means it is included in the encoded subset, and if it corresponds to a “0”, it is excluded in the encoded subset.
            \item Binary increments are independent of the nature or sign of the elements in $S$. This is because binary increments are purely combinatorial and based on inclusion or exclusion of the corresponding elements. Corresponding elements can be in any nature, sign, type of value, abstract symbol/object, repetition, or size, and their combinations can be represented by binary increments of $S$.
            \item Therefore, by generating all possible binary strings of length '$n$' and converting each string into a subset, binary increments ensures that all '$2^n$' possible combinations of elements in $S$ are enumerated.
        \end{enumerate}
    \end{lemma}
    
    However, these enumerations would be computationally challenging as elements in set $S$ increase. Addition of each new element increases the number of combinations exponentially, $2^n$. Despite some subsets equaling to the same number, every number has its unique elements contributing to sum to it.

    
    \subsection{General Applicability Of Patterns Observed}
    Now, in these enumerations of possibilities, there are a number of patterns the research has recognized, but we will focus on the ones important for the method to be presented and later discussions:
    
    \begin{observation}
        The pattern of differences from 0 to any number right before a possibility that is directly an element we took from a set repeats multiple times until the last possibility. E.g.: \{1, 2, 3, 4\}: 

        \begin{itemize}
            \item[] [0, \textbf{(1)}, \textbf{(2)}, 3, \textbf{(3)}, 4, 5, 6], [\textbf{(4)}, 5, 6, 7, 7, 8, 9, 10]
            \item[] [0, \textbf{(1)}, \textbf{(2)}, 3], [\textbf{(3)}, 4, 5, 6], [\textbf{(4)}, 5, 6, 7], [7, 8, 9, 10]
            \item[] [0, \textbf{(1)}], [\textbf{(2)}, 3], [\textbf{(3)}, 4], [5, 6], [\textbf{(4)}, 5], [6, 7], [7, 8], [9, 10]
        \end{itemize}

        Pattern of differences of numbers in each bracket '[\ldots]' are the same through out the possibilities.
    \end{observation}
    
    \begin{observation}
        Starting from both ends of the possibility space and moving inwards, the sum of the 2 values from the opposite sides is equal to the sum of all the elements. This highlights a symmetry in the enumeration. Let us see this: \{1, 2, 3, 4\}: 0, \textbf{(1)}, \textbf{(2)}, 3, \textbf{(3)}, 4, 5, 6, \textbf{(4)}, 5, 6, 7, 7, 8, 9, 10.

        \begin{itemize}
            \item[] \{1 + 2 + 3 + 4\} = 10: 0 + 10 = 10; 9 + 1 = 10; 8 + 2 = 10; 7 + 3 = 10; \ldots
        \end{itemize}
    \end{observation}

    \begin{theorem}
        Observation 1 and 2 are generally applicable to any set's binary increments enumeration.

        \begin{enumerate}
            \item As we established that binary increments focus on the combinations of the corresponding elements of a set rather than the nature of the elements themselves, a binary string means the same thing to any given set in terms of inclusion and exclusion of corresponding elements (Lemma 1).
            \item For example, the binary string '00011' would mean that the subset represented is a combination of the first and second elements from the right corresponding to the bits set to 1.
            \item This would hold true for any set with elements greater than the highest bit position set to 1 (from the left). If the set has lower number of elements than the highest bit position set to 1, then it means the binary string cannot represent the combinations of that set due to insufficient number of elements in the set.
            \item As binary increments are all about combinations, inclusion and exclusion of corresponding elements, binary increments will always start from 000\ldots0 to 111\ldots1. The order of successive inclusion and exclusion remains the same for any given set regardless of the nature of the elements.
            \item In the case of Observation 1. If given any binary string and you separate it into two parts, the lower bits on the right and the higher bits on the left, with each of the two parts having at least one bit of the whole binary string. For any single increment of the higher bits, the lower bits would have to repeat all their combinations. Hence, the pattern of difference (combinations of the lower bits) remains the same and repeats in all the rest of the enumeration (combinations of higher bits), again regardless of the nature of corresponding elements from a set.
            \item In the case of Observation 2. Given any set and you take its binary increment enumeration of possible combinations. The first combination on the left of the enumeration is the empty set, exclusion of all corresponding elements from any set, $000 \ldots 0$. The last combination on the right of the enumeration is the inclusion of all the elements in the set, all bits set to 1, $111 \ldots 1$. Adding these would give you all 1s in the resulting binary string. Moving inwards from both sides of the enumeration you would get an inclusion of the lowest bit from the left/start of the enumeration and an exclusion of the lowest bit from the right/end of the enumeration. Adding these would still result in a binary string of all 1s of '$n$' (number of elements in the set) length. This symmetry of inclusion and exclusion goes on until the middle of the binary increment enumeration of the possibilities of any set, regardless of the nature of the elements in the set.
            \item Therefore, Observation 1 and 2 are applicable to any given set's binary increment enumeration.
        \end{enumerate}
    \end{theorem}
    
    Using Observation 1, we would only need to know one full pattern from 0 to a possibility right before the next element directly from the set, and only a single possibility of the other cycles in order to be able to enumerate all the other possibilities of that cycle. Given set \{1, 2, 3, 4\}, and knowing the first cycle/pattern is 0, \textbf{(1)}, \textbf{(2)}, 3, the starting point of the next cycle we know should be 3. So, the other possibilities would be \textbf{(3)}, 4, 5, 6, and same thing applies for 4; \textbf{(4)}, 5, 6, 7. For the last part of the enumeration, we only know that the ending point of the enumeration is 10, the sum of all the elements in the set, and that will be all the information we need to find the possibilities that come before 10. Apply the pattern from the ending point; \ldots, 7, 8, 9, 10.
    
    As the set gets bigger though, it is not as straightforward because those starting and ending points get further and further apart especially in the far end of the enumeration. When it gets even bigger, the first pattern of 4 possibilities would not reach some possibilities. You might think that increasing the number of possibilities of the first cycle we know to 8 maybe is an answer but, that will not help in a set with 100 elements. It would be clever to implement Observation 2 and find the pattern of possibilities from the end of the enumeration moving inwards, but still, it would not work for even larger sets with larger enumerations. The gap between the starting and ending points would now be vast in the middle of the possibility space rather than at the end of it and still increasing exponentially.
        
    Another great idea would be to spread evenly across the possibility space the starting and ending points we know so that the known cycle would be able to reach any possibility. However, that would be just splitting the exponential growth of possibilities which would also still be infeasible to accomplish for much larger instances. That approach would be like the “Meet in the middle” algorithm\cite{horowitz1974} in terms of splitting the search. All the techniques presented are clever, but they do not escape the exponential growth of possibilities and therefore no polynomial time solution to the problem. This is what most current methods of solving NP-complete problems fall under, clever ways of finding a solution but become exponential in the worst-case.
    
    
    \subsection{*General Applicability And Linear Time And Space Complexity Of Enumerating Only Specific Possibilities}
    Now to unpack the algorithm, we implement Observation 1 at certain points that split the possibility space evenly on each iteration through the elements of the given set. Specifically, we will choose possibilities that split the possibility space into four equal parts on each iteration, using previously enumerated possibilities and new ones derived from them. Let us see this: Given set \{1, 2, 3, 4, 5, 6, 7, 8\}
    
    \begin{enumerate}
        \item We enumerate 0 and the first 2 elements from the set, as we of course know that adding 1 element to the enumeration does not do anything. So, we have: \[0, \textbf{(1)}, \textbf{(2)}, 3, \ldots\]
        \item We move to the next element in the set, enumerate it as a possibility and then add it to certain points to produce other points that would together divide the possibility space into four equal parts. We should add the element on the current iteration from the set to only 3 specific preceding possibilities. Those specific possibilities are, the possibility exactly before the element we previously directly enumerated from the set (Only current bit is set to 1 and all the lower bits except the bit right before the current bit, \ldots101111\ldots), the element we previously enumerated directly from the set (Only current bit and the next lower bit are set to 1, \ldots11000\ldots) and the possibility exactly before the element we currently enumerated directly from the set (Current bit and all lower bits are set to 1 without exception, \ldots1111\ldots). So far, we only have 3 possibilities in the enumeration space after 0 and when we enumerate the next element from the set, it should be added to these preceding possibilities because they are exactly the 3 subsets we describe: \[0, \textbf{(1)}, \textbf{(2)}, 3, \textbf{(3)}, 4, 5, 6, \ldots\]
        \item Next element from the set and add it to the 3 preceding possibilities: \[0, \textbf{(1)}, \textbf{(2)}, 3, \textbf{(3)}, 4, 5, 6, \textbf{(4)}, \_, \_, 7, 7, \_, \_, 10, \ldots\] \[\text{or}\] \[\textbf{0}, (1), (2), \textbf{3}, \textbf{(3)}, 4, 5, \textbf{6}, \textbf{(4)}, \_, \_, \textbf{7}, \textbf{7}, \_, \_, \textbf{10}, \ldots\]
        \item Sums to specific preceding possibilities are placed in their corresponding opposite side of the enumeration. Repeat step 3: \[\textbf{0}, (1), (2), 3, (3), 4, 5, \textbf{6}, \textbf{(4)}, \_, \_, 7, 7, \_, \_, \textbf{10}, \textbf{(5)}, \_, \_, \_, \_, \_, \_, \textbf{11}, \textbf{9}, \_, \_, \_, \_, \_, \_, \textbf{15} \ldots\]
        \item Continue like that for all elements in the set till you get to the last possibility which is the sum of all the elements, 1 + 2 + 3 + 4 + 5 + 6 + 7 + 8 = 36. 
    \end{enumerate}

    \begin{theorem}
        Enumeration of only specific possibilities in a way that splits the possibility space into four equal parts on each iteration through the element of a set can be computed in linear time.

        \begin{enumerate}
            \item We have established that binary increment enumerations can be done for any kind of set regardless of the nature of the elements in the given set (Lemma 1).
            \item Therefore, enumerating only specific positions of the enumeration means the same for any set respective to the sets corresponding elements and order of their correspondence.
            \item Enumerating the first 4 combinations is fairly straightforward and requires constant time and space complexity.
            \item Moving to the enumeration of other specific combinations we see that an addition of a new element from the set gives us 3 more possibilities.
            \item Therefore, for any addition of a new element from the set, we get a total of 4 elements, '$4n$', this is a linear time computation for any number of elements in the given set. Adding the constant time complexity of enumerating the first 4 combinations, the time complexity still remains linear time, \textbf{O$(n)$}.
            \item Same thing applies for space complexity, we would have to compute and store the first four possibilities which is constant time. Plus storing all the specific combinations which grow at a linear rate, \textbf{O$(n)$}, therefore, a linear space complexity.
        \end{enumerate}
    \end{theorem}
        
    From that, we know for any given set, the amount of possibilities we can compute is '$4(n - 2) + 4$'. We subtract from the given set the first 2 elements, then for every other element we get 3 more possibilities and finally with the first 2 elements we have the starting pattern of 4 possibilities hence, $4(n - 2) + 4$, simplified, $4n - 4$. That is the complexity of computing and storing these possibilities, which is a linear function, so the time and space complexity is O$(n)$. As you might have noticed, although this gives us the starting and ending points efficiently, the distance between them still keeps increasing exponentially still.

    Before we tackle that, for the given set \{1, 2, 3, 4, 5, 6, 7, 8\}, let us enumerate all the starting and ending points:
    
    \begin{enumerate}
        \item We start with the first four possibilities, 0, \textbf{(1)}, \textbf{(2)}, 3, \ldots
        \item Then 3 is the next starting point. We add 3 to the possibility exactly before the element we previously directly enumerated from the set which is 1 and get the ending point 4. The distance between this starting and ending \textbf{(S/E) point} is 1, they are next to each other. For consistence, we will use this notation: (starting, ending)\textsubscript{distance}, so, (3, 4)\textsubscript{1}. We now have: 0, \textbf{(1)}, \textbf{(2)}, 3, (\textbf{(3)}, 4)\textsubscript{1}, \ldots
        \item Then we add 3 to the element we previously enumerated directly from the set which is 2 and get the starting point 5. For its ending point, we add 3 to the possibility exactly before the element we currently enumerated directly from the set which is 3 and get the ending point 6. The distance between this S/E point is 1, they are next to each other. Now we have: 0, \textbf{(1)}, \textbf{(2)}, 3, (\textbf{(3)}, 4)\textsubscript{1}, (5, 6)\textsubscript{1}, \ldots
        \item We continue like this adding the current element to the 3 specific preceding possibilities as the possibilities enumerated alternate in being starting and ending points. The rate at which the distance between S/E points increases is '$2^n - 1$' starting from the third element and '$n$' starting from 1. '$n$' increases by 1 per iteration through the elements of a set. All elements always gives us a pair of S/E points except for the first two elements. So, the pair of S/E points we would get from each element would have the same distance between them.
        \item For the set \{1, 2, 3, 4, 5, 6, 7, 8\} we would have in the end, this: 0, \textbf{(1)}, \textbf{(2)}, 3, (\textbf{(3)}, 4)\textsubscript{1}, (5, 6)\textsubscript{1}, (\textbf{(4)}, 7)\textsubscript{3}, (7, 10)\textsubscript{3}, (\textbf{(5)}, 11)\textsubscript{7}, (9, 15)\textsubscript{7}, (\textbf{(6)}, 16)\textsubscript{15}, (11, 21)\textsubscript{15}, (\textbf{(7)}, 22)\textsubscript{31}, (13, 28)\textsubscript{31}, (\textbf{(8)}, 29)\textsubscript{63}, (15, 36)\textsubscript{63}.
    \end{enumerate}
    
    So, to expand the notation, (starting, ending)\textsubscript{distance}, it will be; starting point, \ldots distance minus 1 of unknown possibilities\ldots, ending point. E.g.: (9, 15)\textsubscript{7} = 9, \_, \_, \_, \_, \_, \_, 15.
    
    We have moved up in abstraction and now know how use the notation, (starting, ending)\textsubscript{distance}. Now to tackle the exponential growth in distance between S/E points, we have to notice that Observation 1 can further be implemented in the unknown possibilities between S/E points like so:
    
    \begin{enumerate}
        \item Expand (7, 22)\textsubscript{31}: \[7, \_, \_, \_, \_, \_, \_, \_, \_, \_, \_, \_, \_, \_, \_, \_, \_, \_, \_, \_, \_, \_, \_, \_, \_, \_, \_, \_, \_, \_, \_, 22\]
        \item The expanded pattern from 0 to the 32\textsuperscript{nd} (31 + 1) possibility: \[0, 1, 2, 3, 3, 4, 5, 6, 4, \_, \_, 7, 7, \_, \_, 10, 5, \_, \_, \_, \_, \_, \_, 11, 9, \_, \_, \_, \_, \_, \_, 15\]
        \item Translate it to the (7, 22)\textsubscript{31} S/E point: \[7, 8, 9, 10, 10, 11, 12, 13, 11, \_, \_, 14, 14, \_, \_, 17, 12, \_, \_, \_, \_, \_, \_, 18, 16, \_, \_, \_, \_, \_, \_, 22\] \[(7, 22)_{31} = 7, 8, 9, 10, (10, 11)_1, (12, 13)_1, (11, 14)_3, (14, 17)_3, (12, 18)_7, (16, 22)_7\]
    \end{enumerate}
    
    \begin{corollary}
        (Theorem 2) A zoom-in between any S/E point can also be done in linear time and space complexity.

        \begin{enumerate}
            \item Using Observation 1, the pattern of differences that repeats itself between any S/E point is the pattern of differences between zero and a possibility on a distance of the S/E point away from 0 in the possibility space.
            \item We established that the enumeration of these S/E points can be done in linear time and space complexity (Theorem 2).
            \item Therefore, enumerating the same S/E points between any other S/E points would still take linear time and space complexity.
            \item We would then be enumerating S/E points in a deeper level than the initial S/E point enumeration (within another S/E point). We would need to take note of the S/E point we zoomed into so that when done with its inner S/E points, we go back to the upper level initial S/E point enumeration.
        \end{enumerate}
    \end{corollary}
    
    
    \subsection{*Deepest Depth Level Of Zoom-In Increasing Linear To The Number Of Elements In The Set}
    So far, we can see that this method would function in a First In, Last Out (FILO) or Last In, First Out (LIFO) stack of the S/E points with varying depths. When searching for a possibility, we would zoom-in on different S/E points (finding S/E points within S/E points), moving back when the solution does not exist between a S/E point, and moving to other higher level S/E points. The deepest depth levels would be on the last S/E points and lower depth levels on the first S/E points. This is due to the increase in distance between the S/E points on the far end of the enumeration. The more the distance between the S/E points, the more we would need to zoom-in to reach the deepest possibility.

    \begin{theorem}
        Even in the deepest depth level of zoom-in, the depth level would still be linear to the number of elements in the set.

        \begin{enumerate}
            \item A pattern of difference that would fit in one of the last pair of S/E points, no matter how large/many the possibilities are, is quarter the possibilities of the whole possibility space. This is because of how the method enumerates S/E points, the S/E points are chosen in a way that splits the possibility space into four equal parts. So, the last S/E point pair is together always half the whole possibility space. Exploring any one of them is exploring half of the half possibilities the pair represents, which is quarter of the whole possibility space.
            \item Again, a pattern that would fit into one of the S/E points of the last S/E point pair of those quarter possibilities, would be quarter of the quarter possibilities, making it one sixteenth $\left(\frac{1}{16}\right)$ of the whole possibility space.
            \item Let us look at this in a binary string, say we are given a binary string of 8-bits. Quarter of the possibility space would be represented as, 00111111, one sixteenth would be, 00001111, and further to one sixty fourth $\left(\frac{1}{64}\right)$, 00000011.
            \item Notice that the pattern of bits set to 1 starts from being the third bit from the left and all lower bits, skipping the first 2 highest bits. Then after, it keeps skipping 2 more bits and sets all the lower bits to 1 the deeper we zoom-in. This goes on all up until we get to only the last bit set 1, for odd bit length, and only 2 for even bit length. This means that for an 8-bit binary string (i.e. 8 elements in a set), in the worst case, we would have to zoom-in 4 times, which is half the bit length.
            \item Therefore, for any bit length (number of elements in a set), we would only have to at most zoom-in $\frac{n}{2}$ times to get to any possibility in the possibility space.
        \end{enumerate}
    \end{theorem}
    
    The implementation of Observation 1 can be done for any S/E point and it would give us a zoom-in on the possibilities in between them. From here, we can now get to any possibility in the possibility space of a given set, it would be just a matter of zooming-in on the right S/E points. To choose the right S/E points, there should be a criteria to help us choose the right S/E points. Criteria can differ based on the specific possibility we are looking for. In a Subset Sum Problem, this can be determined using the target sum. An ideal criteria would be a sum-bound criteria that tells us that a target sum exists or not within a S/E point. However, the pattern between the S/E points can be very complex, especially in the mix of negative and positive numbers and can even go out of the sum-bound range that we are given by only the starting and ending points. Let us look at an illustration of how complex the pattern would be between S/E points in the mix of positive and negative numbers, we will use the set \{-12, -7, -3, 5, 8, 10\}: 
    
    \begin{figure}[H]
        \centering
        \includegraphics[width=0.4\textwidth]{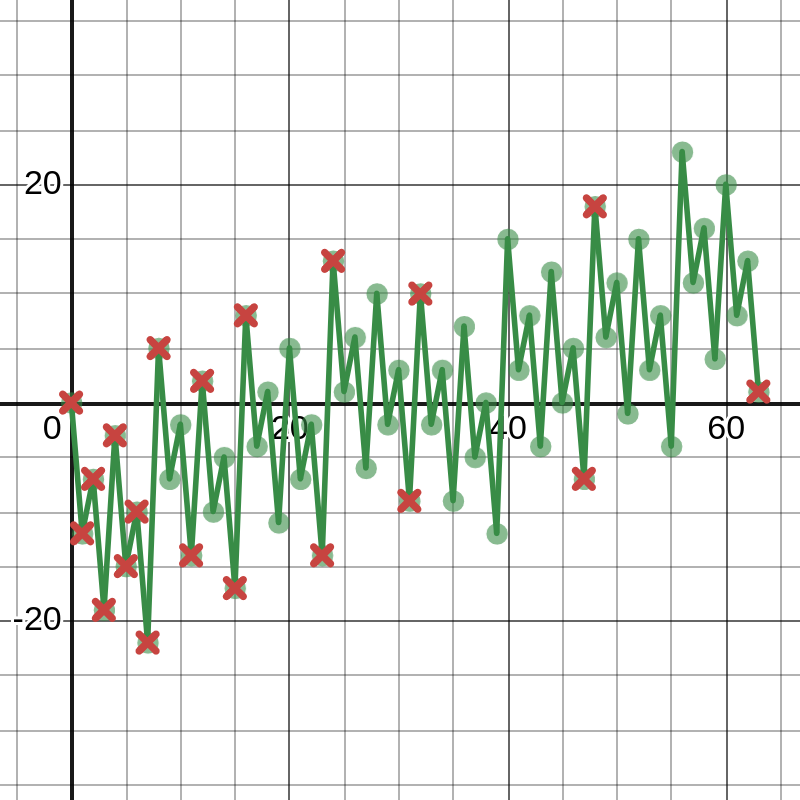}
        \caption{Possibility space plot starting with the negative elements}
    \end{figure}

    \begin{figure}[H]
        \centering
        \includegraphics[width=0.4\textwidth]{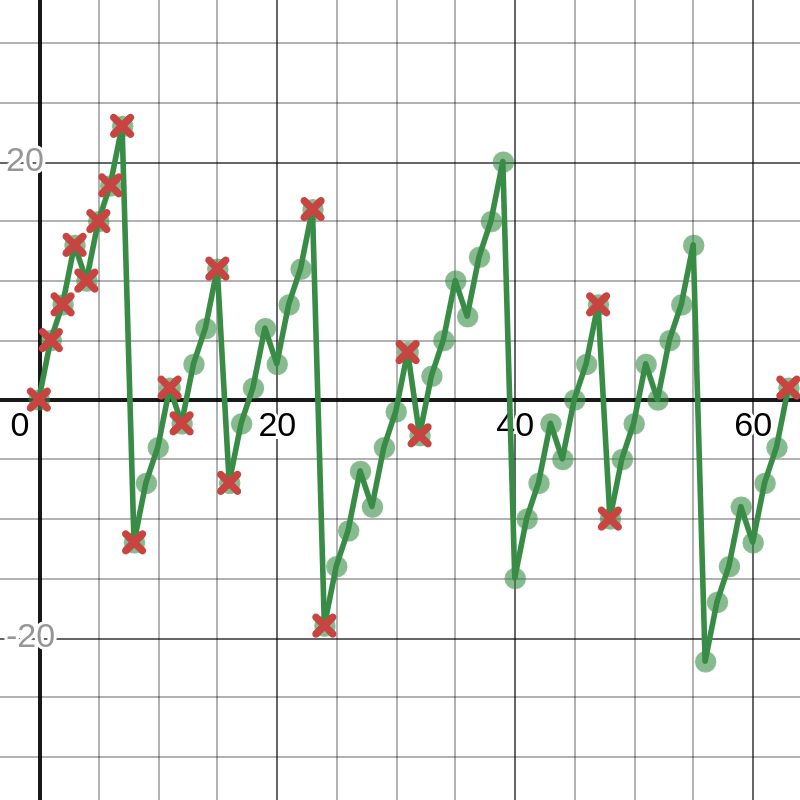}
        \caption{Possibility space plot starting with the positive elements}
    \end{figure}
    
    All circle points are subsets that the set \{-12, -7, -3, 5, 8, 10\} can achieve, and cross points are the specific S/E points of the set. All cross points are also circle points, but not all circle points are cross points. In figure 1, the possibilities are enumerated in binary increments having the negative elements in the lower bits and the positive elements as the higher bits. In contrast, figure 2, has the same possibilities, this time enumerated having positive elements in the lower bits and the negative elements as higher bits. The possibilities in figure 1 exhibit an overall orderly distribution of subsets (less \textbf{global fluctuations}), but the subsets themselves fluctuate a lot (more \textbf{local fluctuations}). In figure 2, the subsets themselves exhibit a more orderly pattern (less local fluctuations) and has minimal fluctuation, but the overall distribution of the subsets is fluctuating a lot (more global fluctuations). Notice also that the last 2 pair of cross points (2 S/E points) have a pattern within them that goes beyond the y-distance (sum-bound) of the 2 pair of cross points in both figures.
    
    A sum-bound criteria for a S/E point would typically be in the form of “\textit{the target sum lies between a starting point that is greater than target sum and ending point less than target sum}”. From these plots, any specific target sum we look for would be a horizontal line equal to the target sum, $y = a$, where '$a$' is the target sum. Then any point in the plot that touches this horizontal line is a possibility/subset of the set that achieves the target sum. If the target sum (horizontal line) is between a S/E point that has more local fluctuations, it would take longer to verify if any of the points in the inner pattern ever touch the target sum. The points of the plot would be crossing the horizontal line multiple times and sometimes not even touch the horizontal line at all. Computing all these checks to sometimes eventually find that none of the points touch the horizontal line can be inefficient. It would be better if the local fluctuations were more of line or with minimal fluctuations, it would then be easy for a sum-bound to perform a single (or few) checks and determine if any of the inner patterns touch the horizontal line. With that, the order of local fluctuations in figure 2 have an advantage over the local fluctuations in figure 1.
    
    A disadvantage for both these plots, figure 1 and 2, is that the sum-bound criteria would fail to correctly determine if the inner points touch a certain horizontal line if some of the inner points go out of the y-distance (sum-bound) of the S/E points. Before that, let us first deduce what causes these local fluctuations in figure 1 and the global fluctuations of subsets in figure 2.
    
    \begin{theorem}
        Ordering the elements of the set \textbf{for each sign} seperately by absolute value/magnitude makes their order of binary increment enumeration not overly fluctuate and have an overall increasing or decreasing pattern. Making it easy for the sum-bound criteria to be applied on S/E points and brings candidates close to each other as possible.

        \begin{enumerate}
            \item In the previous plots, we order the set \{-12, -7, -3, 5, 8, 10\} according to the number line, then either start enumerating the positive or negative elements in the lower bits first then followed by the other sign in higher bits. Order still remains as the number line for the negative elements even when we start with positive elements.
            \item As we change from starting with either the positive or negative elements, the over fluctuations changes from being local fluctuations to being global fluctuations. This signals that one of the order of the elements, positive or negative, is causing this unwanted fluctuation.
            \item Binary increment enumeration of possible combinations for each single-sign set of the elements in the set \{-12, -7, -3, 5, 8, 10\} are: \{-12, -7, -3\} = 0, -12, -7, -19, -3, -15, -10, -22 and \{5, 8, 10\} = 0, 5, 8, 13, 10, 15, 18, 23.
            \item It then becomes apparent that it is the order of the negative elements that causes the erratic fluctuations. Therefore, each sign should be ordered from smallest to biggest according to its magnitude/absolute value. Binary increment enumeration of possible combination for the negative sign set now: \{-3, -7, -12\} = 0, -3, -7, -10, -12, -15, -19, -22.
            \item Like that, the possible combinations for each sign have an overall increasing or decreasing pattern. They can have minimal opposite direction movement, but cannot go beyond the magnitude of the previous element from the set. It can at least be equal though.
            \item Generality of this for any single-sign set is that, if we are given elements \{a, b, c, d, e\}, in binary increments we enumerate '$a$' followed by '$b$' then the sum of '$a$' and '$b$'. The value of '$c$' might be less than the value of '$a + b$', but '$c$' cannot be less than '$b$', it only can at least be equal to '$b$'. Same applies for '$a + b + c$' and '$d$', '$d$' can be less than '$a + b + c$' but cannot be less than '$c$', it only can at least be equal to '$c$'. This way, fluctuations are minimal and are within the sum-bound of S/E points for each sign.
            \item None of the subsets of a set or within a S/E point can even at least be equal to the starting point as they are all additions/translations of preceding subsets to the starting point. This is true as none of the preceding subsets would be equal to zero as zeros are removed from the set before we enumerate S/E points (section 3.9).
            \item None of the subsets of a set or within a S/E point can even at least be equal to the ending point as the ending point is the sum of all element of the set preceded in the enumeration. So, even any translations of these ending points would still be greater in magnitude than all subsets within the S/E point.
            \item Given this ever increasing or decreasing pattern, S/E points would only phase in and phase out once through a target sum (horizontal line). This means all candidates are brought as close to each other as possible.
        \end{enumerate}
    \end{theorem}

    Now let us see how the plot would look like with this new order of elements based on their magnitude/absolute value, \{-3, -7, -12, 5, 8, 10\}:

    \begin{figure}[H]
        \centering
        \includegraphics[width=0.4\textwidth]{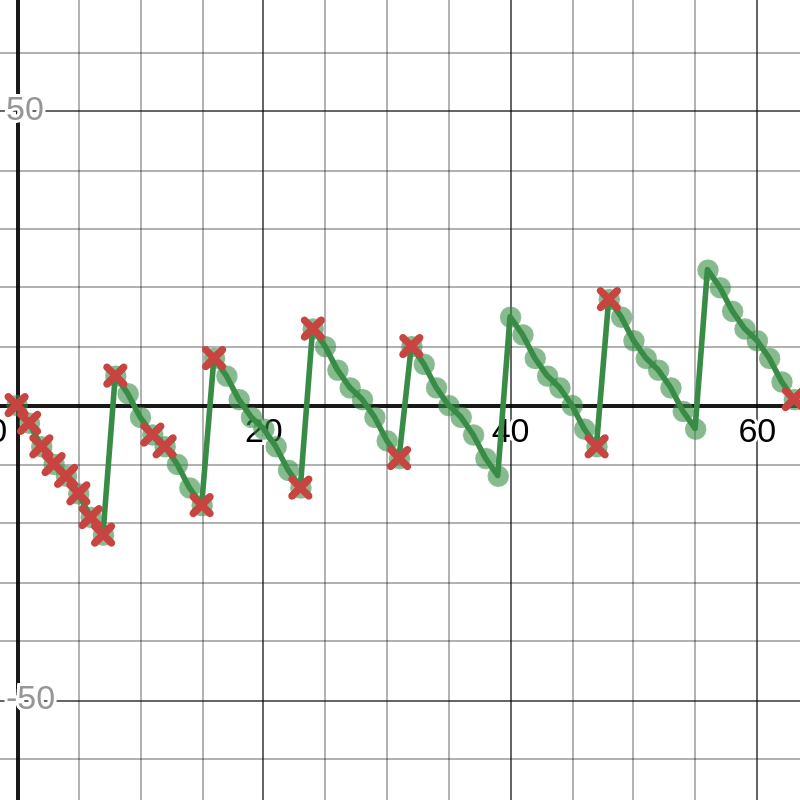}
        \caption{Ordered possibility space plot starting with negative elements}
    \end{figure}

    \begin{figure}[H]
        \centering
        \includegraphics[width=0.4\textwidth]{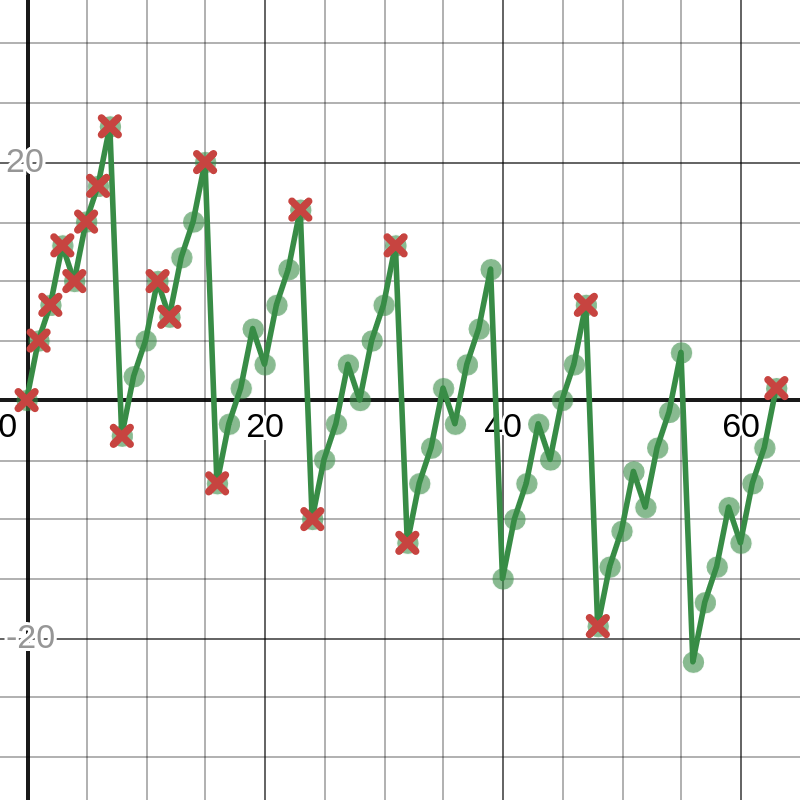}
        \caption{Ordered possibility space plot starting with positive elements}
    \end{figure}
    
    The plots now have a much more overall increase and decreasing pattern both for the local and global fluctuations of the subsets. It would now be easier to apply a sum-bound criteria to these S/E points. However, there is still the problem of the last 2 pair of S/E points (cross points) being exceeded by the inner points, which was a disadvantage for both the plots regardless of whether they start by enumerating the positive or negative elements. Before dealing with that disadvantage, note that for each single-sign set, the S/E points are ever increasing and do not go beyond their sum-bounds given the order we established. Let us also look at how a plot of a single-sign set looks like without mixing the signs. Figure 5 shows a plot of subsets for the set \{1, 2, 3, 4, 5, 6\} and figure 6 shows plots of subsets for the set \{-2, -2, -2, -2, -2, -2\}:
    
    \begin{figure}[H]
        \centering
        \includegraphics[width=0.4\textwidth]{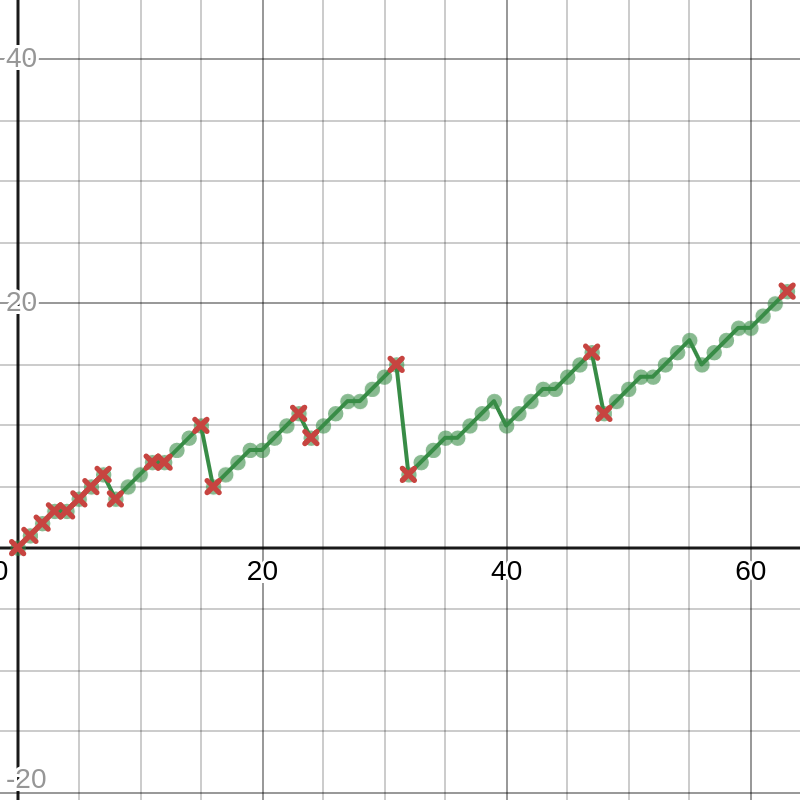}
        \caption{Single-sign set possibility space plot with no repetition of elements}
    \end{figure}

    \begin{figure}[H]
        \centering
        \includegraphics[width=0.4\textwidth]{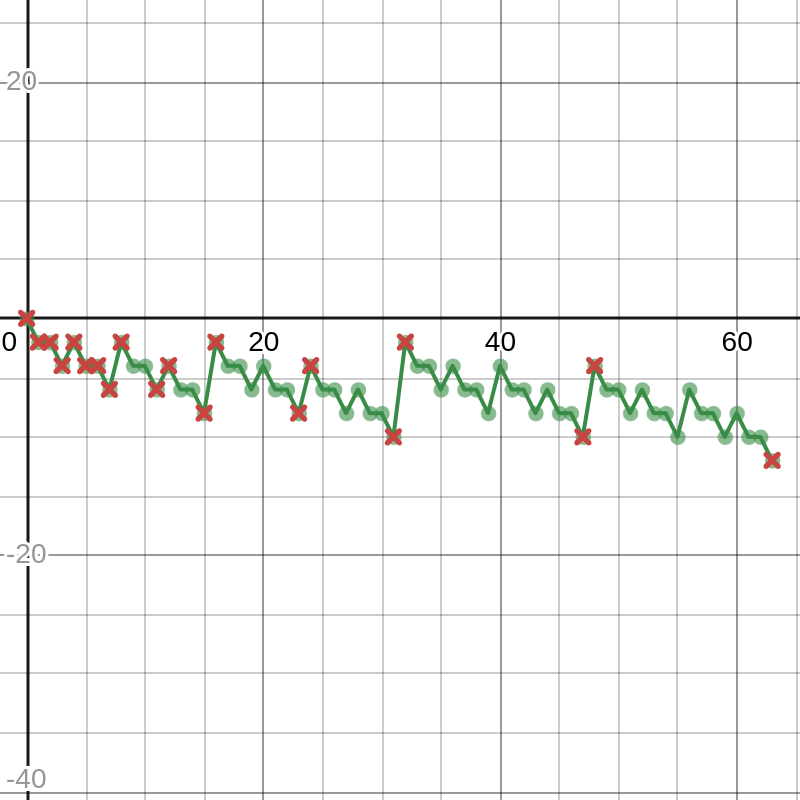}
        \caption{Single-sign set possibility space plot with repetition of elements}
    \end{figure}
    
    In figure 5, we can see that for the set \{1, 2, 3, 4, 5, 6\}, even though there may be fluctuations, the fluctuations are within the ever increasing pattern and do not overly fluctuate. All the change in directions follow what we talked about in terms of the general set \{a, b, c, d, e\}, '$c$' being able to be lower than '$a+b$' but cannot be less than '$b$', can at least be equal to '$b$' (for figure 6). These change in directions are starting points themselves as they are direct elements from the set or translations of them. Therefore, every change in direction is captured as a starting point of some S/E point, at some zoom-in depth level.
    
    Since single-sign sets are most ideal for this method, it would be a good idea to change a mixed-sign set to a single-sign set and still maintain number of subsets and count of unique subsets of the original set in the transformation. All that in the following section.

    
    \subsection{*Transforming A Mixed-Sign Set To A Single-Sign Set}
    The crucial part of this transformation is to maintain the exact number of subsets and count of unique subsets of the original mixed-sign set when transforming it to a single-sign set. We will start by explaining the process of transformation first, then later show some results of the transformation on an example mixed-sign set. The resulting possibilities may have both signs, but the shape of the resulting plot is exactly the same as a shifted single-sign set. This will be apparent when we demonstrate the plot of the resulting shifted elements and S/E points.
    
    \begin{enumerate}
        \item We would first take all the elements and order them all, both negative and positive, not separately, according to magnitude, smallest to biggest.
        \item We would then add up and store the sum of all negative/positive elements (only one sign or the other). Then when reading the elements from the set we would read them as the other sign. As we enumerate the S/E points, we shift them by adding the sum of the negative/positive elements we previously acquired.
        \item Note, we enumerate each point before shifting it, shifting the actual elements of the set first before enumerating the S/E points yields incorrect results.
        \item The resulting set and its S/E point enumeration is the shifted single-sign set.
    \end{enumerate}
    
    Let us go through that using an example and showing its final plot compared to its initial mixed-sign set plot. We will use the set: \{-1, -2, -3, 1, 2, 3\}.
    
    \begin{enumerate}
        \item We order the set according to magnitude for both signs at the same time and not separately: \{-1, 1, -2, 2, -3, 3\}
        \item Adding up either all the positive or negative elements, we will add all the negatives in this case. This gives us a total of '-6'.
        \item Enumerate the S/E points reading all elements as positive since we added together the negative elements in the previous step, and each point we get, we shift by the '-6'. We will start with the S/E points of the set as is, then later shift it with the “-6”. Original S/E points: (0, 1)\textsubscript{1}; (1, 2)\textsubscript{1}; (2, 3)\textsubscript{3}; (3, 4)\textsubscript{3}; (2, 4)\textsubscript{7}; (4, 6)\textsubscript{7}; (3, 7)\textsubscript{15}; (5, 9)\textsubscript{15}; (3, 9)\textsubscript{31}; (6, 12)\textsubscript{31}. Shifted S/E points: (-6, -5)\textsubscript{1}; (-5, -4)\textsubscript{1}; (-4, -3)\textsubscript{3}; (-3, -2)\textsubscript{3}; (-4, -2)\textsubscript{7}; (-2, 0)\textsubscript{7}; (-3, 1)\textsubscript{15}; (-1, 3)\textsubscript{15}; (-3, 3)\textsubscript{31}; (0, 6)\textsubscript{31}.
        \item Finally, the plot of the subsets for both set \{-1, -2, -3, 1, 2, 3\} (figure 7) and \{-1, 1, -2, 2, -3, 3\} (figure 8) shift by '-6' all subsets and read all elements as positive:
    \end{enumerate}
    
    \begin{figure}[H]
        \centering
        \includegraphics[width=0.4\textwidth]{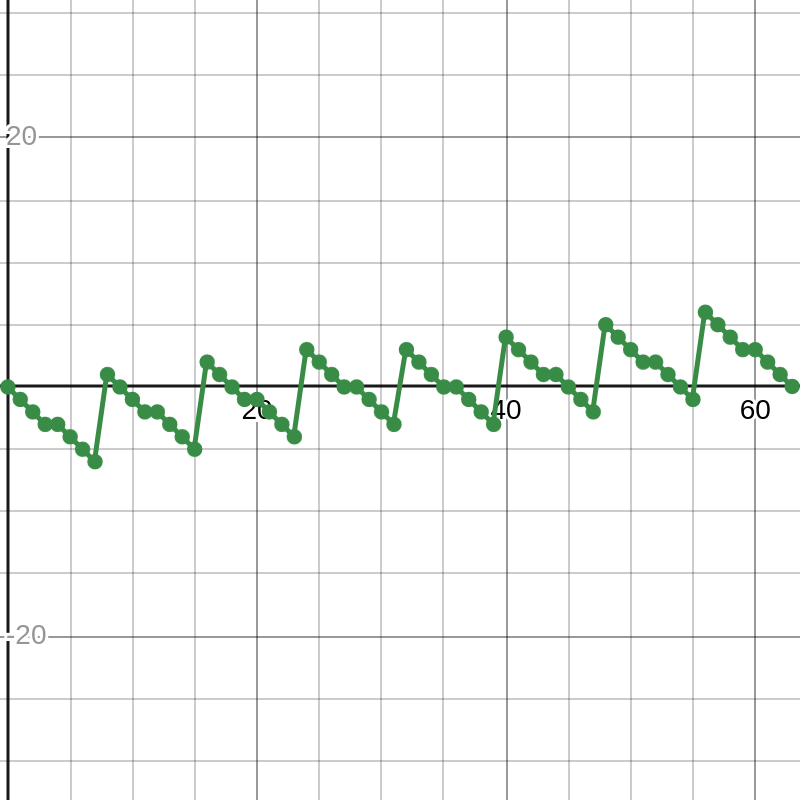}
        \caption{Ordered mixed-sign set possibility space plot}
    \end{figure}

    \begin{figure}[H]
        \centering
        \includegraphics[width=0.4\textwidth]{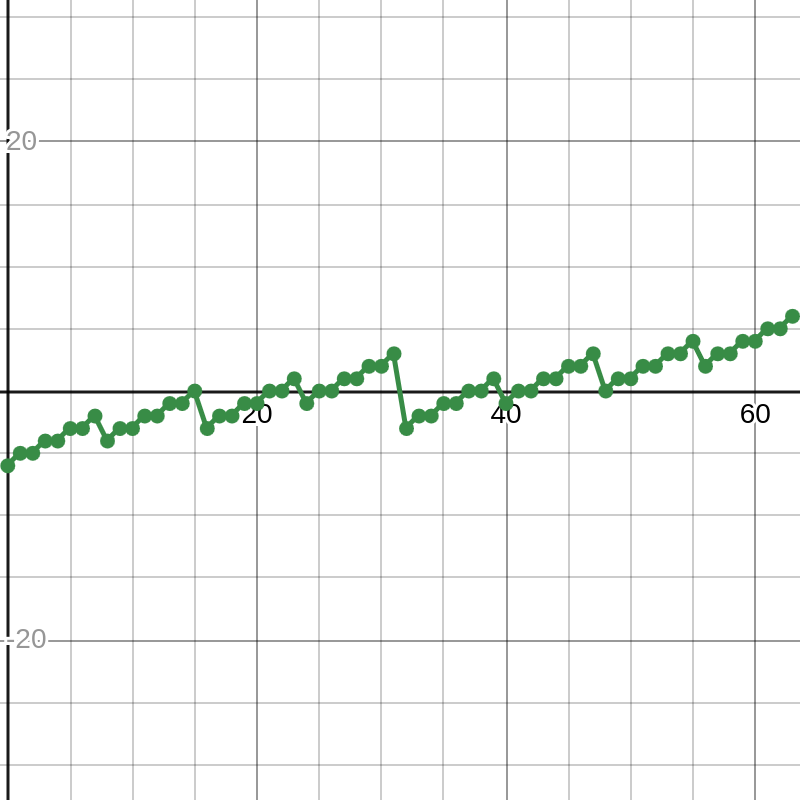}
        \caption{Ordered mixed-sign set plot as a single-sign set possibility space plot}
    \end{figure}

    From these, we can see that the number of subsets and count of unique subsets is the same both for the original set and the transformed set. If we draw the same horizontal line ($y = a$) on both plots, it intersects the same number of points for both figure 7 and 8, this shows the correctness of the transformation. This also tackles the issue of subsets exceeding the sum-bound in mixed sign sets (Theorem 4).
    
    Another thing worth discussing is that, in figure 6, is a sets with repetition and shows a lot of fluctuations. When the method deals with cases as such, there is a lot of redundancy in the subsets explored and hence reducing performance. We still are to discuss these redundancies in detail in section 3.10 and ways of optimizing around these.

    
    \subsection{*General Determination Of Criteria Of S/E Points To Zoom Into}
    \begin{enumerate}
        \item If it is a mixed-sign set, we would first need to transform it as discussed in the previous section. Complexity analysis around this is still also to be dealt with in section 3.13.
        \item We would then always end up with either a single-sign set or a shifted single-sign set, simply just a single-sign set also.
        \item For a single-sign set, you simply have to determine the S/E points for the elements, after ordering them according to their magnitude.
        \item The S/E points would be determined according to the order set on the elements. If it is a negative set, the points would be decreasing and therefore the starting points would be either greater or equal to the target sum and the ending points be less than equal to the target sum. If it is a positive set, the starting points would have to be less than or equal to the target sum and the ending points greater than or equal to the target sum. Otherwise, it would mean that the S/E point does not touch or even cross the target sum (horizontal line). Positive: $S <= T$ and $E >= T$, or Negative: $S >= T$ and $E <= T$.
        \item The time complexity of checking if a S/E point potentially has a solution is constant time complex, \textbf{O$(1)$}. It is a simple comparison of the starting and ending points to the target sum.
        \item In the S/E points, no possibilities within it can be equal to either the starting or ending point, they are all in-between (Theorem 4).
        \item Given the ever increasing or decreasing nature of S/E points and sum-bound, encountering a starting point greater than (in magnitude) the target value we are looking for, disqualifies all the other S/E point to come after, only in that level.
    \end{enumerate}
    
    This way even for a mixed sign set, we would be able to transform it accordingly, and process it to decide solvability. 
    
    For the presence of zeros in a set however, we would not be able to either take it as a positive or negative element, it is neutral. As zeros do not affect the sum of any subset they are added to, but instead just increase the number of ways to uniquely achieve a target sum (or any subsets that can be achieved by a set). We can intuitively explain this as introducing a z-axis in the plot of subsets (figures 1 – 8 above). Adding a zero to a set is like adding another xy-plane of the same plot of subsets on top of another, cause it doubles the number of subsets without affecting their sum. More on this to be discussed in section 3.8.
    
    Now that the method has a sum-bound criteria mechanism, let us look at an example of finding all subsets summing to a target sum for the set \{1, 2, 3, 4, 5, 6, 7, 8\}, say we are looking for target sum 24. A criteria for this set would be that the starting point of any S/E point should be less than or equal to 24, and the ending point to be greater than or equal to 24 since it is a positive signed set. So, ($S <= 24$, $E >= 24$)\textsubscript{x}, '$x$' being any distance between the S/E points. The S/E points of the set \{1, 2, 3, 4, 5, 6, 7, 8\} are: 0, 1, 2, 3, (3, 4)\textsubscript{1}, (5, 6)\textsubscript{1}, (4, 7)\textsubscript{3}, (7, 10)\textsubscript{3}, (5, 11)\textsubscript{7}, (9, 15)\textsubscript{7}, (6, 16)\textsubscript{15}, (11, 21)\textsubscript{15}, (7, 22)\textsubscript{31}, (13, 28)\textsubscript{31}, (8, 29)\textsubscript{63}, (15, 36)\textsubscript{63}. The only S/E points that meet the criteria we established of ($S <= 24$, $E >= 24$)\textsubscript{x} are: (13, 28)\textsubscript{31}, (8, 29)\textsubscript{63}, (15, 36)\textsubscript{63}. To help in determining the position of the 24s if we find them in the end, let us modify the notation a bit. When zooming into a S/E point, we will place the number of possibilities preceding the S/E point in front of it like so: Number of Preceding Possibilities (Starting, Ending)\textsubscript{Distance}. So, for the S/E points that have already met the criteria that we need to zoom into we write: 96(13, 28)\textsubscript{31}, 128(8, 29)\textsubscript{63} and 192(15, 36)\textsubscript{63}. Let us zoom-in into these S/E points and keep on applying the criteria which would lead us to the possibilities of 24, if they exist.
    
    \begin{enumerate}
        \item Translate the first patterns that fits into 96(13, 28)\textsubscript{31}, 128(8, 29)\textsubscript{63}, and 192(15, 36)\textsubscript{63}: \[96(13, 28)_{31} = 13, 14, 15, 16, (16, 17)_1, (18, 19)_1, (17, 20)_3, (20, 23)_3, (18, 24)_7, (22, 28)_7\] \[128(8, 29)_{63} = 8, 9, 10, 11, (11, 12)_1, (13, 14)_1, (12, 15)_3, (15, 18)_3, (13, 19)_7, (17, 23)_7, (14, 24)_{15}, (19, 29)_{15}\] \[192(15, 36)_{63} = 15, 16, 17, 18, (18, 19)_1, (20, 21)_1, (19, 22)_3, (22, 25)_3, (20, 26)_7, (24, 30)_7, (21, 31)_{15},\] \[(26, 36)_{15}\]
        \item S/E points that satisfy the criteria are, 112(18, 24)\textsubscript{7}, 120(22, 28)\textsubscript{7}, 160(14, 24)\textsubscript{15}, 176(19, 29)\textsubscript{15}, 204(22, 25)\textsubscript{3}, 208(20, 26)\textsubscript{7}, 216(24, 30)\textsubscript{7} and 224(21, 31)\textsubscript{15}. Some of these S/E points already have 24s in their starting or ending point, and there is no need to further explore its inner S/E points as none of the inner possibilities can be equal to the starting or ending point (Theorem 4). For those S/E points, we have reached solutions already. S/E point accumulate the number of preceding possibilities accordingly, each from its parent S/E point. Then we further translate the first patterns that fits into these S/E points: \[ 204(22, 25)_3 = 22, 23, 24, 25 \] \[ 120(22, 28)_7 = 22, 23, 24, 25, (25, 26)_1, (27, 28)_1 \] \[ 208(20, 26)_7 = 20, 21, 22, 23, (23, 24)_1, (25, 26)_1 \] \[ 176(19, 29)_{15} = 19, 20, 21, 22, (22, 23)_1, (24, 25)_1, (23, 26)_3, (26, 29)_3 \] \[ 224(21, 31)_{15} = 21, 22, 23, 24, (24, 25)_1, (26, 27)_1, (25, 28)_3, (28, 31)_3 \]
        \item S/E point satisfying the criteria: 184(23, 26)\textsubscript{3}, other S/E points that satisfy the criteria already have a possibility of 24 as their starting or ending point. Translate first pattern that fits into this S/E point: \[ 184(23, 26)_3 = 23, 24, 25, 26 \]
    \end{enumerate}
    
    This tells us that in the set \{1, 2, 3, 4, 5, 6, 7, 8\}, there are 10 subsets that sum to 24, namely, 206(24)\textsubscript{0}, 119(24)\textsubscript{0}, 122(24)\textsubscript{0}, 213(24)\textsubscript{0}, 216(24)\textsubscript{0}, 175(24)\textsubscript{0}, 182(24)\textsubscript{0}, 227(24)\textsubscript{0}, 228(24)\textsubscript{0} and 185(24)\textsubscript{0}. The distance “0” means the number in the bracket is a single possibility and not a range between a S/E point. Then to find elements of the set that contribute to the sum of the 24s, you simply take the number of preceding possibilities of the solution, convert it to binary, and reverse the order of the binary sequence. Primarily, the lowest bit in the binary string should correspond to the element we start with when enumerating S/E points, and so on up to the highest bit corresponding to the last element we consider when enumerating S/E points. Compare the binary sequence to the set, and for every element of the set that corresponds to a “1” means that it contributes to the sum of that 24. Take 206(24)\textsubscript{0}, convert 206 to binary you get “11001110”, in reverse, “01110011”, then finally compare this binary sequence to the set: \[ 1, 2, 3, 4, 5, 6, 7, 8 \] \[ 0, 1, 1, 1, 0, 0, 1, 1 \] \[ \text{Therefore: } 2 + 3 + 4 + 7 + 8 = 24 \]
    
    For mixed-sign set instances, the resulting binary string from the preceding possibilities of the solutions would have to be processed further. In the binary string, we would have to negate bits that correspond to elements we added together and shifted by, reading them as the opposite sign. Otherwise, we take the bit as it is. The resulting binary string then would correctly identify which elements correspond to add up to the solution even for a set that is given as a mixed-sign set.
    
    Now, the method is completely explained and we know how to get to any specific possibility in the possibility space. We can describe the possibility space of any set explored by the method as different levels and aisles of possibilities that we can navigate guided by a criteria. That is the algorithmic structure of the Subset Sum Problem. Computation of each level/depth as a whole can be done in linear time and space. Note, the '$4n - 4$' space complexity, is the auxiliary space (space used by the processes of the method/algorithm). This is separate from the input space (space used to store basic information of a subset sum problem such as the set and target sum). The distinction between the two is important because we still have to further optimize the performance of the method. For a decision version of the Subset Sum Problem, the method runs through the possibility space and will find the first possible solution as soon as it can.
    
    Considering that the auxiliary space complexity of '$4n - 4$' on each level of zoom-in might be inefficient in some practical cases, we can further optimize this and make the auxiliary space complexity remain constant on each level. A '$4n - 4$' time complexity would not be impractical though, thanks to the advancement of modern day processors.

    
    \subsection{*Reducing Auxiliary Space Complexity Of Computing S/E Points On Each Level Of The LIFO Stack From linear To Constant}
    \begin{enumerate}
        \item To determine the criteria of S/E points to zoom into for single-sign sets or shifted single-sign sets, we simply check the sign of the elements or change to one, then either apply the $S <= T$ and $E >= T$ or $S >= T$ and $E <= T$ criteria.
        \item We would then need only 6 variables to successively move across all the S/E points of one level. One variable would be a utility variable, 4 variables would be 2 S/E points (as each S/E point represents 2 know values, 2 of the 4 variables for each S/E point), and one for storing the distance between each of the S/E points pairs per element.
        \item If dealing with the decision variant of the Subset Sum, we might not need the number of preceding possibilities as we just want to determine if a solution exists or not. However, if needed, including the number of preceding possibilities would introduce 1 new variable to be managed.
        \item Initially, the 4 variables that store 2 S/E points would store the first 4 possibilities of a set. The utility variable would be initialized to 0 but overwritten by the element on each iteration through the elements of a set from the input storage. Let us name the variables $A$ to $E$, $E$ being the utility variable. If $A$ to $D$ contains the first four possibilities, the variables are updated for the third element by assigning the third element to $A$ and translating variables $B$ to $D$ by $A$. Otherwise, the 4 variables would be updated and overwritten on each iteration as will follow. Appropriate shifts have to be applied using the opposite sign numbers total and the starting point of the set, which is the empty set zero, or starting points of S/E point we are zooming into (Section 3.15):
        \begin{enumerate}
            \item $E$ is assigned the current element from the set
            \item Update $B$: $B = (D - A) + E$
            \item Update $C$: $C = A + E$
            \item Update $D$: $D = D + E$
            \item Update $A$: $A = E$
        \end{enumerate}
        \item This would update the variables accordingly and successively move through the S/E points of each level. After each iteration, $A$ is the starting point and $B$ is the ending point for S/E point 1. $C$ is the starting point and $D$ is the ending point for S/E point 2.
        \item Therefore, all this would be simple arithmetic operations performed on only 6 (7 at most) variables, making the space complexity of computing S/E points on a single level of the LIFO stack constant.
    \end{enumerate}
    
    
    \subsection{Role Of Repeated Insertion Of Zeros In A Set}
    \begin{enumerate}
        \item Given a set of only repeated zeros $n$ times, all combinations that can be made out of this set always results in a zero as zero does not affect the sum when it is added to any combination.
        \item As established that binary increments enumerate all combinations of any set and the number of combinations is exponential to the number of elements in the set. Therefore, the number of ways we can combine a set of only zeros is $2^n$, where $n$ is the number of the zeros in the set.
        \item Regardless of any other non-zero element added to the set containing repeated zeros, this still does not affect the number of ways we can uniquely combine the zeros without the inclusion of other non-zero elements in the set.
        \item For a set containing only a repetition of a single non-zero element, combinations that can be made from the set to achieve a target sum equal to the repeated element is exactly equal to the number of repeated elements in the set. This is because in order to achieve the target sum equal to the repeated element, an element would be taken as is and not combined with any other element. If combined with another element, it would be much greater (if repeated element is positive) or much smaller (if repeated element is negative) than the target sum.
        \item Repeated non-zero elements would have a distribution curve that has peak at the repeated element, then for greater or lesser sums, the number of subsets that achieve those sums would taper off and decrease in frequency. This is because to achieve the greater or lesser sums, more of those repeated elements are used up.
        \item Instead, in a set containing repeated zeros and other non-zero elements, inserting even one more zero affects even the number of ways to achieve any other non-zero target sum. This is because even the non-zero subsets can be added with a zero (as a single element) or subsets summing to zero (combination of the zeros) without affecting their overall sum. The number of ways to achieve the non-zero target sum would be '$k \times 2^n$', where $k$ is the number of ways to achieve the non-zero target sum. Adding even a single zero increases '$n$' and therefore increases the exponent in '$k \times 2^n$'.
        \item As intuitively previously explained, this would be the same as placing layers of the same possibility space (xy-plane) plots on top of each other where the number of layers increase exponentially with respect to the number of zeros in the set. This creates what we term “\textbf{z-axis exponentiation}”.
        \item Therefore, any insertion of non-zero elements (repeated or not) in a set does not make the number of ways to achieve a specific target sum increase exponentially in the z-axis. Only a repeated insertion of zero causes the number of ways to achieve any specific target sum to exhibit an exponential z-axis increase rate.
        \item All other non-zero elements would only affect the rate of increase of subsets equaling a specific target sum only on the xy-plane. One case of high increase rate of subsets (only on the xy-plane) equaling the same target sum is in sets with repetition. Their plot have the most fluctuations than any other plot. More on this in section 3.10.
    \end{enumerate}

    
    \subsection{*Overhead For The Presence Of Zeros In A Set And Its Complexity}
    \begin{enumerate}
        \item The overhead would have to count and not pass through to the method any zeros (this does not affect the zero of the empty set in the binary increment enumeration). This way the main factor causing z-axis exponential growth is out of the way.
        \item The overhead should function as a filter as the elements of the set are entered, then only one check would be made for each element. If it is a non-zero, the element is passed to the method, but if it is a zero, the overhead would just increment the count of zeros and not pass anything to the method.
        \item The time complexity of this overhead would be linear to '$n$' as each element is checked only once and the space complexity is constant as the overhead would only need to store the count of zeros, incrementing it each time a zero is encountered in the elements.
        \item The overhead can also take on the task of adding either all the negative or positive elements for mixed-sign set instances. So, one more variable to hold the total of either all the positive or negative elements. The variable will be accessed later to shift the S/E points of the now to be read as all positive or negative set.
    \end{enumerate}
    
    
    \subsection{*Optimization For Sets With Repetition And More Practical Optimizations}
    Let us see how the redundancy in subsets explored by the method occurs in sets with repetition. We will look at the S/E points of a general set with repetition and observe patterns that could hint redundancy. In the following, each variable represents a repeated element:
    
    For set \{$k$, $k$, $k$, $k$, $k$, $t$, $t$, $t$, \ldots\}, S/E points are these: 0, $k$, $k$, $2k$, $(k, 2k)_{1}$; $(2k, 3k)_{1}$; $(k, 3k)_{3}$; $(2k, 4k)_{3}$; $(k, 4k)_{7}$; $(2k, 5k)_{7}$; $(t, t+4k)_{15}$; $(t+k, t+5k)_{15}$; $(t, t+5k)_{31}$; $(2t, 2t+5k)_{31}$; $(t, 2t+5k)_{63}$; $(2t, 3t+5k)_{63}$; $(t, 3t+5k)_{127}$; $(2t, 4t+5k)_{127}$; \ldots
    
    The set can contain any number of elements that may repeating any number of times without restriction. As long as the elements are ordered by absolute value and all read as one sign in case of a mixed sign set (Sections 3.4 and 3.5). This makes sure the repeating elements come consecutively. No matter how many times an element repeats, there is one universal observation we made for repeating elements:
    \begin{itemize}
        \item \textbf{Observation}: We choose to explore the first 4 possibilities/subsets fully then start checking for starting point similarities after them. Let us see the full subsets of every S/E point pair produced by repeating elements '$k$' and '$t$'. First, for '$k$':
        \begin{enumerate}
            \item \textbf{$(k, 2k)$\textsubscript{1}; $(2k, 3k)$\textsubscript{1} = $k$, $2k$, $2k$, $3k$.}
            \item \textbf{$(k, 3k)$\textsubscript{3}; $(2k, 4k)$\textsubscript{3} = \{$k$, $2k$, $2k$, $3k$\}, $2k$, $3k$, $3k$, $4k$.}
            \item \textbf{$(k, 4k)$\textsubscript{7}; $(2k, 5k)$\textsubscript{7} = \{$k$, $2k$, $2k$, $3k$, $2k$, $3k$, $3k$, $4k$\}, \{$2k$, $3k$, $3k$, $4k$\}, $3k$, $4k$, $4k$, $5k$.}
        \end{enumerate}
        Now, for '$t$':
        \begin{enumerate}
            \item \textbf{$(t, t+4k)$\textsubscript{15}; $(t+k, t+5k)$\textsubscript{15} = $t$, $t+k$, $t+k$, $t+2k$, $t+k$, $t+2k$, $t+2k$, $t+3k$, $t+k$, $t+2k$, $t+2k$, $t+3k$, $t+2k$, $t+3k$, $t+3k$, $t+4k$, $t+k$, $t+2k$, $t+2k$, $t+3k$, $t+2k$, $t+3k$, $t+3k$, $t+4k$, $t+2k$, $t+3k$, $t+3k$, $t+4k$, $t+3k$, $t+4k$, $t+4k$, $t+5k$.}
            \item \textbf{$(t, t+5k)$\textsubscript{31}; $(2t, 2t+5k)$\textsubscript{31} = \{$t$, $t+k$, $t+k$, $t+2k$, $t+k$, $t+2k$, $t+2k$, $t+3k$, $t+k$, $t+2k$, $t+2k$, $t+3k$, $t+2k$, $t+3k$, $t+3k$, $t+4k$, $t+k$, $t+2k$, $t+2k$, $t+3k$, $t+2k$, $t+3k$, $t+3k$, $t+4k$, $t+2k$, $t+3k$, $t+3k$, $t+4k$, $t+3k$, $t+4k$, $t+4k$, $t+5k$\}, $2t$, $2t+k$, $2t+k$, $2t+2k$, $2t+k$, $2t+2k$, $2t+2k$, $2t+3k$, $2t+k$, $2t+2k$, $2t+2k$, $2t+3k$, $2t+2k$, $2t+3k$, $2t+3k$, $2t+4k$, $2t+k$, $2t+2k$, $2t+2k$, $2t+3k$, $2t+2k$, $2t+3k$, $2t+3k$, $2t+4k$, $2t+2k$, $2t+3k$, $2t+3k$, $2t+4k$, $2t+3k$, $2t+4k$, $2t+4k$, $2t+5k$.}
            \item \textbf{$(t, 2t+5k)$\textsubscript{63}; $(2t, 3t+5k)$\textsubscript{63} = \{$t$, $t+k$, $t+k$, $t+2k$, $t+k$, $t+2k$, $t+2k$, $t+3k$, $t+k$, $t+2k$, $t+2k$, $t+3k$, $t+2k$, $t+3k$, $t+3k$, $t+4k$, $t+k$, $t+2k$, $t+2k$, $t+3k$, $t+2k$, $t+3k$, $t+3k$, $t+4k$, $t+2k$, $t+3k$, $t+3k$, $t+4k$, $t+3k$, $t+4k$, $t+4k$, $t+5k$, $2t$, $2t+k$, $2t+k$, $2t+2k$, $2t+k$, $2t+2k$, $2t+2k$, $2t+3k$, $2t+k$, $2t+2k$, $2t+2k$, $2t+3k$, $2t+2k$, $2t+3k$, $2t+3k$, $2t+4k$, $2t+k$, $2t+2k$, $2t+2k$, $2t+3k$, $2t+2k$, $2t+3k$, $2t+3k$, $2t+4k$, $2t+2k$, $2t+3k$, $2t+3k$, $2t+4k$, $2t+3k$, $2t+4k$, $2t+4k$, $2t+5k$\}, \{$2t$, $2t+k$, $2t+k$, $2t+2k$, $2t+k$, $2t+2k$, $2t+2k$, $2t+3k$, $2t+k$, $2t+2k$, $2t+2k$, $2t+3k$, $2t+2k$, $2t+3k$, $2t+3k$, $2t+4k$, $2t+k$, $2t+2k$, $2t+2k$, $2t+3k$, $2t+2k$, $2t+3k$, $2t+3k$, $2t+4k$, $2t+2k$, $2t+3k$, $2t+3k$, $2t+4k$, $2t+3k$, $2t+4k$, $2t+4k$, $2t+5k$\}, $3t$, $3t+k$, $3t+k$, $3t+2k$, $3t+k$, $3t+2k$, $3t+2k$, $3t+3k$, $3t+k$, $3t+2k$, $3t+2k$, $3t+3k$, $3t+2k$, $3t+3k$, $3t+3k$, $3t+4k$, $3t+k$, $3t+2k$, $3t+2k$, $3t+3k$, $3t+2k$, $3t+3k$, $3t+3k$, $3t+4k$, $3t+2k$, $3t+3k$, $3t+3k$, $3t+4k$, $3t+3k$, $3t+4k$, $3t+4k$, $3t+5k$.}
            \item \ldots
        \end{enumerate}
        \item[] From the subsets above, the only new, non-redundant subsets on each S/E point pair are subsets not segmented/parenthesized by '\{\ldots\}'. They are the only subsets that we would have to actually explore. The other segmented subsets are repetitions and partial repetitions of previous subsets and should be trivially explored as they are redundant. Arithmetic expression representations of the segmented subsets is as will follow, showing how we observed the subsets repeat and partially repeat. As we are concerned about the count of solutions within the redundant and non-redundant subsets, we add the expressions representing each segmented and unsegmented subsets. We denote the unsegmented subsets as '$A$', meaning the actual exploration solutions count in the non-redundant subsets for each S/E point pair. Note, '$A$' is an actual solutions count that represent different subsets for every S/E point pair, hence we do not perform any arithmetic operation on '$A$' outside of its S/E point pair. Instead, we refer to the S/E point pair as a whole representing all its subsets including its '$A$' abstracted away. For visual simplicity, let us also assign variables to each S/E point pair, $(t, t+4k)_{15}$; $(t+k, t+5k)_{15}$ = '$x$', $(t, t+5k)_{31}$; $(2t, 2t+5k)_{31}$ = '$y$', $(t, 2t+5k)_{63}$; $(2t, 3t+5k)_{63}$ = '$z$'. Let us also continue the S/E points with 4 more '$t$' elements, meaning 4 more S/E point pairs and assign variables to them also, $(t, 3t+5k)_{127}$; $(2t, 4t+5k)_{127}$ = '$a$' \& $(t, 4t+5k)_{255}$; $(2t, 5t+5k)_{255}$ = '$b$' \& $(t, 5t+5k)_{511}$; $(2t, 6t+5k)_{511}$ = '$c$' \& $(t, 6t+5k)_{1023}$; $(2t, 7t+5k)_{1023}$ = '$d$.
        \begin{enumerate}
            \item \textbf{$(t, t+4k)$\textsubscript{15}; $(t+k, t+5k)$\textsubscript{15} = $A$}.
            \item \textbf{$(t, t+5k)$\textsubscript{31}; $(2t, 2t+5k)$\textsubscript{31} = \{$x$\} + $A$}.
            \item \textbf{$(t, 2t+5k)$\textsubscript{63}; $(2t, 3t+5k)$\textsubscript{63} = \{$y$\} + \{$y - x$\} + $A$}.
            \item \textbf{$(t, 3t+5k)$\textsubscript{127}; $(2t, 4t+5k)$\textsubscript{127} = \{$z$\} + \{$z - y$\} + \{$z - y - (y - x)$\} + $A$}.
            \item \textbf{$(t, 4t+5k)$\textsubscript{255}; $(2t, 5t+5k)$\textsubscript{255} = \{$a$\} + \{$a - z$\} + \{$a - z - (z - y)$\} + \{$a - z - (z - y) - (z - y - (y - x$))\} + $A$}.
            \item \textbf{$(t, 5t+5k)$\textsubscript{511}; $(2t, 6t+5k)$\textsubscript{511} = \{$b$\} + \{$b - a$\} + \{$b - a - (a - z)$\} + \{$b - a - (a - z) - (a - z - (z - y))$\} + \{$b - a - (a - z) - (a - z - (y - z)) - (a - z - (z - y) - (z - y - (y - x)))$\} + $A$}
            \item \textbf{$(t, 6t+5k)$\textsubscript{1023}; $(2t, 7t+5k)$\textsubscript{1023} = \{$c$\} + \{$c - b$\} + \{$c - b - (b - a)$\} + \{$c - b - (b - a) - (b - a - (a - z))$\} + \{$c - b - (b - a) - (b - a - (a - z)) - (b - a - (a - z) - (a - z - (z - y)))$\} + \{$c - b - (b - a) - (b - a - (a - z)) - (b - a - (a - z) - (a - z - (z - y))) - (b - a - (a - z) - (a - z - (z - y)) - (a - z - (z - y) - (z - y - (y - x))))$\} + $A$}
            \item \ldots
        \end{enumerate}
        \item[] These expressions simply show how we observed how the preceding subsets subtract each other to produce repeating and partially repeating subsets in the current S/E point pair before the non-redundant subsets come. We do not have a closed formula to how they subtract each other, but simply that the first segment is exactly the previous S/E point pair subsets; the second segment equal to the previous S/E point pair subsets minus subsets in S/E point pair before it and so on. However, an interesting observation unveils when we work out the brackets in those expressions and simplify:
        \begin{enumerate}
            \item \textbf{$(t, t+4k)$\textsubscript{15}; $(t+k, t+5k)$\textsubscript{15} = $A$}.
            \item \textbf{$(t, t+5k)$\textsubscript{31}; $(2t, 2t+5k)$\textsubscript{31} = $x + A$}.
            \item \textbf{$(t, 2t+5k)$\textsubscript{63}; $(2t, 3t+5k)$\textsubscript{63} = $2y - x + A$}.
            \item \textbf{$(t, 3t+5k)$\textsubscript{127}; $(2t, 4t+5k)$\textsubscript{127} = $3z - 3y + x + A$}.
            \item \textbf{$(t, 4t+5k)$\textsubscript{255}; $(2t, 5t+5k)$\textsubscript{255} = $4a - 6z + 4y - x + A$}
            \item \textbf{$(t, 5t+5k)$\textsubscript{511}; $(2t, 6t+5k)$\textsubscript{511} = $5b - 10a + 10z - 5y + x + A$}.
            \item \textbf{$(t, 6t+5k)$\textsubscript{1023}; $(2t, 7t+5k)$\textsubscript{1023} = $6c - 15b + 20a - 15z + 6y - x + A$}.
            \item \ldots
        \end{enumerate}
        \item[] Notice, the coefficents resemble the pascal triangle. 
        \item[] \textbf{Optimization}: 
        \item[] The method would have to only explore the '$A$' subsets when there is repetition and skip over redundant subsets in $$\frac{n^2 + n}{2}$$ steps, where '$n$' is the number of times an element is repeated, excluding the first appearance of the element. The size of '$A$' remains the same as the number of subsets when the repeating element first appeared as long as there is repetition. So, if there is repetition, the number of subsets we actually explore during repetition remains the same. If there is no solution found in any of the '$A$' subsets during the repetition, there would also be zero trivial solutions we skipped. If we have to count the trivial solutions given that at least one of the '$A$' subsets has a count greater than zero, we would have to start from the first repetition adding the trivial counts and multiplying them with the correct coefficents. Remember that each '$A$' subsets are abstracted for each S/E point pair, and before counting later trivial solutions, the previous S/E point pair/s should be added all up together (all trivial counts multiplied by coefficents and added together plus the S/E point pairs' '$A$'). Notice also how the arithmetic operations '+/-' alternate in the trivial subsets' expressions except for '$A$' that is always added. A formula for generating the coefficents of each S/E point pair is: $$p \left( \frac{n - k}{k} \right)$$ where '$p$' is the previous coefficent but starts as 1. '$n$' is the number of times an element is repeated, including first appearance of the element, and '$k$' is the coefficent number we want to generate starting from 1. The number of coefficents in each level is equal to the number of times an element is repeated, excluding first appearance of the element. The count of coefficents are for trivial subsets only, not the '$A$' subsets.
    \end{itemize}
    
    More practical optimizations over other trivial instances. First, the \textbf{Pure Zero Instance}, this is a Subset Sum instance where the target sum is zero and the set contains only zeros. You can simply answer the decision variant of this instance by proposing any of the elements in the set as a solution but, let us go an extra mile to the counting variant. By simply implementing the '$k \times 2^n$' from section 3.8 and the section 3.9's overhead, the method would increment '$n$' as it is the count of zeros entered into the program. It would not pass anything to be actually processed by the method meaning, '$k$' will not be increased at all and will remain as zero. As a result, at the end of computation the program will output that it found zero solutions equaling to the target sum '0'. '$0 \times 2^n$', where '$n > 0$', this expression will always give us zero. Solution, a variable should be added to '$k$', a pure zero instance constant, that will be '1' when the instance is a pure zero instance and '0' when it is not. We denote the pure zero instance constant as '$p$', and all together, '$(k + p) \times 2^n$'. Now, if it is a pure zero instance, '$p$' turns to '1' and even if '$k$' becomes '0', the expression simplifies to '$2^n$'.

    Second, an all even numbers set having an odd number target sum instance. No matter how many even numbers you add up, you can never get an odd number. The expression '$2k$' represents an even number for any real number you enter in '$k$', and '$2k + 1$' represents an odd number for any real number you enter in '$k$'. So, adding up any number of terms '$2k$', results in an expression that is divisible by '2', meaning the sum of all even numbers always result in an even number and can not reach an odd target sum. Adding up odd numbers on the other hand, '$2k + 1$', results in numbers that alternate between being odd and even. This is because the '+ 1' term does also cumulatively add up and result to expressions divisible by '2'. Therefore, given only even number elements in the set and an odd target sum, there is no solution to the instance.
    
    
    \subsection{Generalization Of The Algorithm To The NP Class}
    As we are developing an algorithm for the Subset Sum Problem, an NP-complete problem, it only comes natural to ask how the algorithm generalizes to other NP/NP-complete problems. This section aims to outline a way we suggest of reducing NP class problems to Subset Sum in potential practical complexities than previously thought.

    It was proven by Stephan Cook in 1971 that every problem in the NP class can be reduced to the SAT problem in polynomial time\cite{cook1971}. Reduction from NP problems to SAT as shown by Cook involves encoding to SAT the computation of a Nondeterministic Turing Machine (NTM) that can solve the original NP problem in polynomial time. The satisfiability of the resulting SAT translates to a solution to the original NP problem, and unsatisfiability translates to the original NP problem having no solution. The NTM is a theoretical abstraction concept whose computation we do not know. Its computation is abstracted behind the assumption that “it can solve the NP problem in polynomial time”. The size of the resulting SAT after reduction is then proven to be increasing polynomially to any original NP problem size. This proved SAT as the first NP-complete problem. A problem that is both in the NP class and all NP problems reducible to it. Other NP problems, including the Subset Sum problem, were later also proven to be NP-complete by efficiently reducing known NP-complete problem to them. This implied that any problem in NP can also be efficiently reduced to Subset Sum.

    That is theoretically sound, but in practice, naively encoding to SAT the diverse constraints of NP problems plus the chain of reductions leading to a target problem can have polynomial complexities of high degrees. The high degrees make the practical implementation of the reduction framework too slow for moderate to large instances. Some NP problems are easier to reduce by naive encoding to others, and some are not. This is due to the diverse structures and constraints of NP problems. If the problems have totally distant constraints and structure, reducing one to the other can be complex. On the other hand, some NP problems are naturally reducible to others with little to no adjustments. Therefore, tailored and constraint aware reductions function much more efficient in practice. A more practical reduction framework centering around the Subset Sum we suggest is as follows:

    We encode only the NP/NP-complete problems' structural components to a form of "set of elements and target sum/condition". Then, creatively translate the constraints of the original problem to the structural components now represented as "set of elements and target sum/condition". This is way, the reduction complexity is only in the encoding of the structural components of the original NP problem to Subset Sum. This might provide reductions more efficient than previously thought compared to reductions that worry about encoding constraints. If we standardize how we encode structural components of each problem and correctly translate the different constraints, then any NP problem coming with similar structure and constraints can be reduced easily. Standardized reductions can then be reused and learned from. This looks like quite an actionable plan, in the next section we provide a reduction of $k$-SAT and CNF-SAT to Subset Sum that functions in linear time. The reduction is purely encoding of structural components of CNF-SAT to Subset Sum, then original SAT problem constraints are creatively implemented on the reduced structural components.
    
    
    \subsection{Reduction Of k-SAT And General Conjunctive Normal Form (CNF) SAT To The Subset Sum Problem}
    $k$-SAT is a boolean formula consisting of Conjunctions ($\wedge$) of clauses that have a fixed number of '$k$' disjunctioned ($\vee$) literals per clause. General CNF-SAT on the other hand is a "hybrid $k$-SAT". The clauses in CNF-SAT are not restricted to all have '$k$' literals, they come in any number of literals per clause. We will start off with a simple 3-SAT instance and then build up to the applicability of the method to $k$-SAT and general CNF-SAT. Let us reduce a 3-SAT problem with 4 variables and 4 clauses to demonstrate how the method works through the section: $$(x_1 \vee \neg x_2 \vee x_3) \wedge (\neg x_1 \vee x_2 \vee \neg x_4) \wedge (x_2 \vee \neg x_3 \vee x_4) \wedge (\neg x_1 \vee x_3 \vee \neg x_4)$$

    For every variable we can assign a truth value true or false. We need to determine a truth assign into the 4 variables that will make the whole SAT formula satisfied or evaluate to 'true'. In order for the whole SAT formula to be true, all clauses should be true. First, we need to map all literals in the SAT formula, including negation of variables, to decimal (\textbf{Literal Mapping}): $$x_1 = 1$$ $$\neg x_1 = -2$$ $$x_2 = 4$$ $$\neg x_2 = -8$$ $$x_3 = 16$$ $$\neg x_3 = -32$$ $$x_4 = 64$$ $$\neg x_4 = -128$$

    We alternate the sign in the literal mappings to represent the negation of variables and making it easier to also identify algorithmically that a variable negation is represented. We then now have to determine the value of each clause (\textbf{Clause Value}) by adding its literals according to the mapping of literals we have just done. $$(x_1 \vee \neg x_2 \vee x_3) = 1 - 8 + 16 = 9$$ $$(\neg x_1 \vee x_2 \vee \neg x_4) = -2 + 4 - 128 = -126$$ $$(x_2 \vee \neg x_3 \vee x_4) = 4 - 32 + 64 = 36$$ $$(\neg x_1 \vee x_3 \vee \neg x_4) = -2 + 16 - 128 = -114$$

    The reduction is then complete and ready for the Subset Sum Solver. It is not entirely the work of the Subset Sum Solver that helps us find solutions for SAT instances, but a work around the set of literal mappings and of clause values we already have, \{1, -2, 4, -8, 16, -32, 64, -128\} and \{9, -126, 36, -114\} respectively. The literal mapping set as a set in the Subset Sum and clause value set as a set of target conditions/sums. Before detailing that, let us analyze the complexity of this reduction.

    \subsubsection{General Reduction Of $k$-SAT and general CNF-SAT To Subset Sum And Its Complexity}
    \begin{enumerate}
        \item Mapping variables/literals and their negations in the SAT formula into decimals. Alternate the sign in powers of 2 in the literal mapping to make it easier to algorithmically determine if the represented literal is a negation. Computing this is linear time complex to the number of literals in the formula.
        \item Next, we get all the value of the clauses, this is achieved by adding all the literal mappings of each clause. Doing this for each clause is linear time complex to the number of clauses. The number of literals to add up to get the clause values is at worst the number of literals in the biggest clause (i.e. for general CNF-SAT), for $k$-SAT the number of literals to add up is the same throughout the formula.
        \item The reduction is then complete and ready for the Subset Sum solver. Adding all those complexities, the reduction is overall linear time complex. \textbf{O$(n + m)$}, where '$n$' is the number of literals and '$m$' is the number of clauses in the given SAT instance.
    \end{enumerate}

    Now, to detail how we make sure that we accurately represent the constraints of the original SAT and how we work around the literal mappings and clause values to get exact solutions:

    \begin{lemma}
        For each clause, we can make combinations of '$n$' literals in a clause in '$2^n$' ways, where '$n$' is the number of literals in the clause.

        \begin{enumerate}
            \item For each literal, we either take the literal or its negation. So, 2 choices per literal in the clause.
            \item To get the number of ways we can combine the literals in a clause we have to multiply the 2 choices for each variable together exactly the number of literals times.
            \item Therefore, the combinations we can make in a clause with its variables is '$2^n$'.
        \end{enumerate}
    \end{lemma}

    \begin{lemma}
        Out of the '$2^n$' combinations of literals in a clause, there are pairs of clause combinations throughout the '$2^n$' combinations that contradict when they are both present in a SAT instance.

        \begin{enumerate}
            \item Every combination of literals in a clause has a complete opposite pair clause such that, completely satisfying one of the clauses (satisfying all its literals), dissatisfies the other.
            \item This would happen because in a pair of clauses that are complete opposites, when one clause takes a variable, the other clause takes the negation of the variable, and vice versa for all the literals in the clause.
            \item Therefore, if such a pair exists in a SAT formula, assignments we make to those clauses should be in a way that would not completely satisfy one clause and consequently dissatisfy the opposite clause.
        \end{enumerate}
    \end{lemma}

    \begin{lemma}
        Complete satisfaction (satisfying all literals) of one clause in the '$2^n$' combinations can satisfy all other clause combinations except for only the complete opposite clause.
        
        \begin{enumerate}
            \item By definition of a complete opposite clause in Lemma 3, an opposite clause is the only clause that takes an negation of a variable when its opposite clause takes a variable and vice versea.
            \item This means that, all other clauses will miss being a complete opposite clause by some literals or at least by one literal.
            \item Therefore, completely satisfying one clause combination can satisfy all clauses having the same literals and number of literals ('$2^n$' combinations), except for only the complete opposite clause. No complete satisfaction of one, and no dissatisfaction of its pair for all clause combinations that are not the complete opposite clause.
        \end{enumerate}
    \end{lemma}

    To avoid ambiguity, we will refer to these clauses talked about from Lemma 2 to 4 as \textbf{Root Clauses}. This can be any clause you pick with '$k$' literals. Lemma 2 to 4 speaks to only the combinations of literals you can make using only the literals in the root clause you picked. \textbf{Clause Signature} refers to and identifies a clause by the specific literals (or its negation) and the number of literals in the clause. Combinations of literals in a root clause include all clauses having the same signature as the picked root clause. These different combinations of literals in the root clauses can either be present or absent in a SAT formula. The presence and absence of the root clauses affects satisfiability of the formula as we will discuss. Therefore, there are '$2^n$' ways of combining the literals of a root clause, meaning there are '$2^n$' clauses having the same signature as any root clause we choose, inclusively (Lemma 2). Signatures of the root clause always have pairs that completely contradict when they are both present in a SAT formula (Lemma 3). Completely satisfying any root clause signature can satisfy all signatures of the root clause except for its complete opposite root clause signature (Lemma 4).

    \begin{lemma}
        All other clauses (non-Root Clauses) in a SAT formula would use some or none of the literals in the root clauses.

        \begin{enumerate}
            \item All other non-Root Clauses can use combinations of other literals and literals in the root clauses or, none of the literals in the root clauses in their clauses.
            \item This is applicable to any arbitrary non-Root Clause size (in terms of number of literals).
        \end{enumerate}
    \end{lemma}

    \begin{lemma}
        A "powers of 2" Subset Sum instance always has one solution for any achievable target sum to its instance, and no false positives.

        \begin{enumerate}
            \item Given a set of ordered elements that are powers of 2 starting from 1, the binary increments enumeration of this set gives a simple numbers count from 0 to the sum of all elements.
            \item There is no repetition and no skipped number between 0 and the sum of all the elements in the achievable subsets.
            \item Given set \{1, 2, 4, 8, 16, 32, 64, \ldots\}, binary increments enumeration is as follows: 0, \textbf{(1)}, \textbf{(2)}, 3, \textbf{(4)}, 5, 6, 7, \textbf{(8)}, 9, 10, 11, 12, 13, 14, 15, \textbf{(16)}, 17, 19, 20, 21, 22, 23, 24, 25, 26, 27, 28, 29, 30, 31, \ldots.
            \item Remember also that when reading the literal mappings set, even though it is a mixed-sign set, we enumerate its S/E points reading the elements of the set as one sign. We only later shift S/E points using the opposite sign elements total. So, even if it is a mixed-sign set as we did for the literal mapping, there is still one solution to any achievable target sum after transformation and shifts.
        \end{enumerate}
    \end{lemma}

    Using those lemmas, a chain of implications and forced assignments guide us to satisfying assignments. We would love to cover the concept extensively, but to make the paper not too lengthy, we will detail only an outline of how the lemmas can be deterministically used to eventually reach satisfying assignments. We will also discuss the general applicability of the method in $k$-SAT and general CNF-SAT problems. All deeper complexity analysis and research results will be done in a follow up paper.

    \subsubsection{General Method For Solving $k$-SAT And General CNF-SAT}
    \begin{enumerate}
        \item Pick any clause in the formula as one of your root clauses.
        \item By picking one root clause, any clause in the formula have the same signature as the root clause you picked, is also a root clause.
        \item Now, we would have to check the presence and absence of contradicting clause pairs with the root clause signature in the formula. This is the part where the Subset Sum Solver helps us algorithmically work back the literals in a clause using the clause values set.
        \begin{enumerate}
            \item If there is a contradicting clause pair in the formula, the 2 clauses are not considered as a clause we have to fully satisfy and we ignore the pair in terms of giving us the first lead to solutions (Lemma 3).
            \item If all the contradicting root clause pairs are included in the formula, the formula is simply not satisfiable.
            \item If there is a root clause signature in the formula that does not have a contradicting root clause pair in the formula, we have to completely satisfy that clause (satisfy all its literals). The complete satisfaction of that clause gives us the first lead/hint of the satisfying assignment solution. For the contradicting pair not in the formula, we should not completely satisfy it as that would completely dissatisfy its pair in the formula (Lemma 4).
            \item If we have a pair that is not in the formula, we can complete satisfy any of those clauses. Completely satisfying them does not dissatisfy any root clause in the formula. Complete satisfaction of both/any of the 2 clauses is a viable first hint/lead to a satisfying assignment solution.
        \end{enumerate}
        \item We then would have to run through the clauses using the satisfying assignment lead from the previous step. For every clause in the formula, we assign truth values we currently know, and leave out what we do not. When assigning the truth values we know in the different clauses, the clauses enforce implication of what should be assigned to other literals based on the truth values we currently have.
        \begin{enumerate}
            \item If our assignment to clause literals makes even one literal evaluate to "true", we do not satisfy or assign to any of the literals in that clause.
            \item If our assignment to clause literals makes all literals we assign to evaluate to "false" but we end with more than one literal we do not have an assignment to, we do not satisfy or assign anything to those literals.
            \item If our assignment to clause literals makes all literals evaluate to "false" and leaves only one literal unassigned, the literal MUST be satisfied. The value we assign to the literal to satisfy it is the value we take as an assignment that should be made for that variable in the formula based on the truth values we have.
            \item All the clauses that do not force us to make any assignments to its literals will have to be revisited. All clauses forcing us to assign a truth value, can be skipped and not revisited.
            \item If we reach a point where no clauses enforce an assignment to any literals we still do not have assignments to, this mean that solutions exist in multiple directions we can take. Here, we would have to make a "true" or "false" assignment to any one of the variables we do not have values for. We assume one direction and see where it further takes us. We run through the clauses we should again and this leads to further implications and forced assignments to literals. That is, if there are clauses that have not been satisfied by our assignment, if all clauses are already satisfied though, solutions exist in all directions we can take. In that case, the number of solutions would be $2^n$, where '$n$' is the number of literals we do not have assignments for.
            \item If at any point the truth values we assign and currently know dissatisfy a clause, this mean that we ran into a false positive and we would have to change the root clause lead assignment completely. 
        \end{enumerate}
        \item Using those rules, they would lead us to determining if a SAT instance is satisfiable or not. If not satisfiable, we would have a root clause having all its signatures in the formula or all root clause assignments leading to false positives. If it is satisfiable, we can reach the exact satisfying assignment solutions.
        \item That is the algorithmic structure of the SAT problem and a lot goes unsaid. Does the method work better when the clause values are ordered? Do other clauses help us decide solvability faster if they are root clauses compared to others? All that to be detailed in a follow up paper including complexity analysis and research results.
    \end{enumerate}

    Let us implement the outlined method to the 4 variable, 4 clause 3-SAT instance we had in the beginning of the section: $$(x_1 \vee \neg x_2 \vee x_3) \wedge (\neg x_1 \vee x_2 \vee \neg x_4) \wedge (x_2 \vee \neg x_3 \vee x_4) \wedge (\neg x_1 \vee x_3 \vee \neg x_4)$$

    \begin{enumerate}
        \item We already have the literal mappings and clause values sets, \{1, -2, 4, -8, 16, -32, 64, -128\} and \{9, -126, 36, -114\} respectively.
        \item We pick a root clause in the clause values: $9 = (x_1 \vee \neg x_2 \vee x_3)$. Clause signatures of this root clause are: 

        \begin{table}[H]
            \centering
            \begin{tabular}{cccc}
                \underline{$(x_1 \vee x_2 \vee x_3)$} & \underline{$(x_1 \vee x_2 \vee \neg x_3)$} & $\overline{(x_1 \vee \neg x_2 \vee x_3)}$ & \underline{$(x_1 \vee \neg x_2 \vee \neg x_3)$} \\ 
                \underline{$(\neg x_1 \vee \neg x_2 \vee \neg x_3)$} & \underline{$(\neg x_1 \vee \neg x_2 \vee x_3)$} & \underline{$\cancel{(\neg x_1 \vee x_2 \vee \neg x_3)}$} & \underline{$(\neg x_1 \vee x_2 \vee x_3)$}
            \end{tabular}
        \end{table}

        Every root clause signature has its complete opposite pair above/below it. An underline of a clause means the clause is not part of the formula, and an overline means the clause is part of the formula. We call any solutions we get from an overlined root clause signature "\textbf{Internal Solution}" because, the root clause is part of the formula. We call any solutions we get from an underlined root clause signature "\textbf{External Solutions}" because, the root clause is not part of the formula. The \underline{$\cancel{(\neg x_1 \vee x_2 \vee \neg x_3)}$} clause is cancelled out because we must not completely satisfy it. Completely satisfying it would dissatisfy its pair in the formula, $\overline{(x_1 \vee \neg x_2 \vee x_3)}$. We do not want to dissatisfy any clause in the formula. All other clause signatures can be completely satisfied as their pairs are not part of the formula. Complete satisfaction of any one of the other signatures can satisfy the root clause signature in the formula, $\overline{(x_1 \vee \neg x_2 \vee x_3)}$, including the complete satisfaction of the clause itself.
        \item Complete satisfaction of each of those root clause signatures, excpet for the one cancelled out, is as will follow. We will represent truth value assignments to variables $x_1$ to $x_4$ as: ****. '*' = unknown assignment, 'T' = assign true and 'F' = assign false. '\textsubscript{I}****' before a truth value assignment, represents that the solution is from an internal solution root clause, and '\textsubscript{E}****', means the solutions represented are from an external solution root clause. $$_{E}TTT*$$ $$_{E}TTF*$$ $$_{I}TFT*$$ $$_{E}TFF*$$ $$_{E}FFF*$$ $$_{E}FFT*$$ $$_{E}FTT*$$
        \item Moving through the clauses the impliactions on those external and internal solution assignments we have is as follows:
        \begin{enumerate}
            \item First Clause: $(x_1 \vee \neg x_2 \vee x_3)$ \[ _{E}TTT* = (T \vee F \vee T)\text{; Satisfies its felow root clause. No forced assignments. } =\ _{E}TTT* \] \[ _{E}TTF* = (T \vee F \vee F)\text{; Satisfies its felow root clause. No forced assignments. } =\ _{E}TTF* \] \[ _{I}TFT* = (T \vee T \vee T)\text{; Root clause satisfies itself. No forced assignments. } =\ _{I}TFT* \] \[ _{E}TFF* = (T \vee T \vee F)\text{; Satisfies its felow root clause. No forced assignments. } =\ _{E}TFF* \] \[ _{E}FFF* = (F \vee T \vee F)\text{; Satisfies its felow root clause. No forced assignments. } =\ _{E}FFF* \] \[ _{E}FFT* = (F \vee T \vee T)\text{; Satisfies its felow root clause. No forced assignments. } =\ _{E}FFT* \] \[ _{E}FTT* = (F \vee F \vee T)\text{; Satisfies its felow root clause. No forced assignments. } =\ _{E}FTT* \]
            \item Second Clause: $(\neg x_1 \vee x_2 \vee \neg x_4)$ \[ _{E}TTT* = (F \vee T \vee *)\text{; Clause evaluates to true. No forced assignments. } =\ _{E}TTT* \] \[ _{E}TTF* = (F \vee T \vee *)\text{; Clause evaluates to true. No forced assignments. } =\ _{E}TTF* \] \[ _{I}TFT* = (F \vee F \vee *)\text{; Clause forced assignment. } x_4 = F. =\ _{I}TFTF \] \[ _{E}TFF* = (F \vee F \vee *)\text{; Clause forced assignment. } x_4 = F. =\ _{E}TFFF \] \[ _{E}FFF* = (T \vee F \vee *)\text{; Clause evaluates to true. No forced assignments. } =\ _{E}FFF* \] \[ _{E}FFT* = (T \vee F \vee *)\text{; Clause evaluates to true. No forced assignments. } =\ _{E}FFT* \] \[ _{E}FTT* = (T \vee F \vee *)\text{; Clause evaluates to true. No forced assignments. } =\ _{E}FTT* \]
            \item Third Clause: $(x_2 \vee \neg x_3 \vee x_4)$ \[ _{E}TTT* = (T \vee F \vee *)\text{; Clause evaluates to true. No forced assignments. } =\ _{E}TTT* \] \[ _{E}TTF* = (T \vee T \vee *)\text{; Clause evaluates to true. No forced assignments. } =\ _{E}TTF* \] \[ _{I}TFTF = (F \vee F \vee F)\text{; False positive. Clause Dissatisfied. } =\ \cancel{_{I}TFTF} \] \[ _{E}TFFF = (F \vee T \vee F)\text{; Clause evaluates to true. Assignment satisfies clause. } =\ _{E}TFFF \] \[ _{E}FFF* = (F \vee T \vee *)\text{; Clause evaluates to true. No forced assignments. } =\ _{E}FFF* \] \[ _{E}FFT* = (F \vee F \vee *)\text{; Clause forced assignment. } x_4 = T. =\ _{E}FFTT \] \[ _{E}FTT* = (T \vee F \vee *)\text{; Clause evaluates to true. No forced assignments. } =\ _{E}FTT* \]
            \item Fourth Clause: $(\neg x_1 \vee x_3 \vee \neg x_4)$ \[ _{E}TTT* = (F \vee T \vee *)\text{; Clause evaluates to true. No forced assignments. } =\ _{E}TTT* \] \[ _{E}TTF* = (F \vee F \vee *)\text{; Clause forced assignment. } x_4 = F. =\ _{E}TTFF \] \[ _{E}TFFF = (F \vee F \vee T)\text{; Clause evaluates to true. Assignment satisfies clause. } =\ _{E}TFFF \] \[ _{E}FFF* = (T \vee F \vee *)\text{; Clause evaluates to true. No forced assignments. } =\ _{E}FFF* \] \[ _{E}FFTT = (T \vee T \vee F)\text{; Clause evaluates to true. Assignment satisfies clause. } =\ _{E}FFTT \] \[ _{E}FTT* = (T \vee T \vee *)\text{; Clause evaluates to true. No forced assignments. } =\ _{E}FTT* \]
        \end{enumerate}
        \item Therefore, all the satisfying assignment solutions to the given instance are: $_{E}TTT*$, $_{E}TTFF$, $_{E}TFFF$, $_{E}FFF*$, $_{E}FFTT$, $_{E}FTT*$. '*' means solutions exist in multiple directions we can take. All these satisfy the formula, including those with unassigned value to literal $x_4$. So, for the assignments having '*', solutions exist in all directions. There is at most one literal left unassigned in these solutions, so, those would count as 2 solutions ($2^n$, where equal 1 for the literal that has no assignment). The given instance has 9 satisfying assignment solutions.
    \end{enumerate}
    
    
    \subsection{Overall Time And Space Complexity Analysis}
    In this section we synthesize into one all the components of the method (tagged with asterisks '*' in their headings) and analyze the complexity of the method as a whole.
    
    \begin{enumerate}
        \item \textbf{Time Complexity:}
            \begin{itemize}
                \item[] \textbf{Auxiliary Time Complexity (only):}
                \item Given any arbitrary set of any size, we can compute its first level S/E points efficiently. Specifically, we only need to compute the first 4 possibilities using the first 2 elements from the set, then for every other element in the set we compute 3 more possibilities. Therefore, for any set we compute a linear amount of S/E points, \textbf{O$(n)$} (Theorem 2).
                \item Using the sum-bound criteria, the method is able to check each S/E point if a solution might exist within or not, in constant time, \textbf{O$(1)$} (section 3.6). It is only when the sum-bound criteria hints a potential solution within a S/E point we zoom-in and further perform other constant time checks. We successively move through the S/E points of each level linearly with respect to the number of possibilities/elements in the set or fitting within that S/E point. This is done all up to the lowest level of S/E points with distance 1 and it would be a matter of 1 of the points touching the target value or none (Theorem 4).
                \item Given the order we set on the elements in the set before they are processed for searching and the binary increment order of enumerating the possibility space, which we navigate by S/E points, a possibility we are looking for cannot randomly exist anywhere in the possibility space (Theorem 4). There are S/E points where the possibility is out of bound (below or above the solution) and S/E points where solutions potentially exists (in-bound), false positives may be encountered. When S/E points start to be above bound, there is no need to continue exploring S/E points of that level. What increases the number of S/E points where solutions might exist (or Candidates) are fluctuations in the possibility space plots like the ones in figures 1 – 8. The plot increases the number of fluctuations based on the density of elements in the set (difference between elements).
                \item A horizontal line (target sum) passing through a region of the plot having more fluctuations means more S/E points being hinted to have solutions. Since we have optimized for repetitions in section 3.10, the method is capable of handling repetition and trivially passing over redundant areas of the possibility space. Other instances with variation in the elements in the set have a possibility space plot that is more stretched out away from the x-axis than the one with repetition. That means that a horizontal line (target sum) intersects the plot much lesser, meaning lesser S/E points satisfying the criteria or being hinted as potentially having solutions than plots representing repetition.
                \item From the first level of the LIFO stack, S/E points with the highest distance we can zoom-in into are the last 2 S/E points. The method has to use all the elements in the set to get to the last 2 S/E points because it successively moves throught the possibility space. To further get to the last 2 inner S/E points of those first level largest S/E points, we would have to use 2 less the number of elements in the set, and so on reducing 2 elements per zoom-in until we can not zoom-in further (section 3.4, Theorem 3). So, if the method has to always zoom-in on the largest S/E point on every level to reach a solution the time complexity would be: \[ \textbf{O}(n) + \textbf{O}(n - 2) + \textbf{O}(n - 4) + \ldots \]
                \item[] \begin{center} This has an arithmetic series so, we can simply use the arithmetic series formula: \end{center} \[ \textbf{O}\left( \frac{d}{2} (a_1 + a_d) \right) \]
                \item[] \begin{center} '$d$' denotes the zoom-in depth for '$n$' elements and '$d = \frac{n}{2}$', '$a_1$' is the first term and '$a_d$' is the last term.\end{center} \[ \textbf{O}\left( \frac{d}{2} (n + (n - 2(d - 1))) \right)\] \[ \textbf{O}\left( \frac{n}{4} \left(n + \left(n - 2\left( \frac{n}{2} - 1 \right)\right)\right) \right) \] \[ \textbf{O}\left( \frac{n}{4} (n + (n - n + 2)) \right) \] \[ \textbf{O}\left( \frac{n}{4} (n + 2) \right) \] \[ \textbf{O}\left( \frac{n^2 + 2n}{4} \right) \] \[ \textbf{O}\left( n^2 \right) \]
                \item[] Therefore, that is the time complexity to reach the furthest possibility in the possibility space for this algorithm. In other words, that is the time complexity of the algorithm to travel its longest path from start to end, independent of other path complexities.
                \item Traversing other paths (start to end) would independantly have complexity costs less than the complexity above. Lowest complexity starts from sub-linear, not using all elements in the set to reach a solution.
                \item To traverse paths during runtime however, we do not need to start over a path to chance to it. The are main and sub-branches. It takes changing only a sub-branch to reach the next \textbf{candidate}: true solution or false positive at the end of another path. All traversal is guided by the sum-bound criteria. Therefore, the time complexity of reaching the first candidate is less than or equal to \textbf{O$(n^2)$}. Moving from candidate to candidate after reaching the first candidate, is under the \textbf{O$(n^2)$} complexity.
                \item The positioning, count and distribution of true solutions and false positives in the possibility space is governed by the complex combinatorics of the elements in the set, their density, instance size, and target sum. This makes closed form complexities hard to formulate. However, given that reaching the first candidate and moving candidate to candidate is less than or equal to \textbf{O$(n^2)$}, the behaviour of the algorithm on different cases is as follows:
                \item \textbf{Polynomial or less size of the in-bound solution space}. When the total number of true solutions and false positives is polynomial or less relative to the instance size, the algorithm:
                \begin{enumerate}
                    \item Can efficiently decide solvability of the  instance. Finding at least one solution.
                    \item Can efficiently decide unsolvability of the instance. Going through all the false positives and verify that none of them lead to solutions.
                    \item Can efficiently count all solutions to the instance if they exist.
                \end{enumerate}
                \item \textbf{Exponential size of the in-bound solution space}. When the total number of false positives and true solutions is exponential* relative to the instance size, the algorithm:
                \begin{enumerate}
                    \item Can efficiently decide solvability only if reaching the first true solution is not buried by an exponential number of false positives.
                    \item Cannot efficiently decide unsolvability as it would have to go through an exponential number of false positives and verify that they all do not lead to solutions.
                    \item Cannot efficiently count all solutions but, can at least efficiently keep listing solutions if they exist as candidate to candidate complexity is less than quadratic time.
                \end{enumerate}
                \item *The total number of candidates (false positives and true solutions) can not be exactly exponential ($2^n$) relative to the instance size unless only if the instance is a Pure Zero Instance. All subsets in the possibility space would have to be equal to the same value for the number of candidates to be exactly equal to '$2^n$'. Otherwise in other instances, the number of candidates would be sub-exponential in the worst case.
                \item Trivial solvability (Pure Zero Instance, even target sum in instances with only repetition, etc.) and unsolvability (All even elements and odd target sum, out of bounds target sum, etc.) are also handled appropriately (Sections 3.6 and 3.10).
                \item The sparser the elements in the set, the easier it is for the method to decide solvability and unsolvability. Conversely, the denser the elements in the set, the harder it is to decide solvability and unsolvability. Beyond a certain threshold of density though, the number of solutions starts increasing exponentially. In extreme instances such as elements that are consecutive numbers from 1, there are only solutions and no false positives. Deciding solvability then becomes easy and trivial as we can even efficiently keep on listing more than one solution.
                \item Putting all these auxiliary complexities together we get a worst case time complexity \textbf{O$(C)$}, where '$C$' is the number of candidates for the specific instance and is sub-exponential. The average and best case complexities can be \textbf{O$(C)$}, where '$C$' is polynomial or less.
                \item[] \textbf{Including Input Storage Manipulation Time Complexity}:
                \item About the overhead that filters zeros (section 3.9), as mentioned, it would function as a filter and would check every element only once as each element is entered, checking if it is a zero or not. A non-zero element is passed on and a zero element is not passed but just increments the count of zeros. This functions exactly linear to the  number of elements. \textbf{O$(n)$}.
                \item The overhead also has to either choose whether the final set is negative or positive shaped when it is given as a mixed-sign set. Either way, we get the correct number of subsets and the count of unique subsets represented. Even if the overhead chooses to be one sign while the set has more elements of the other sign, processing them is simply cumulatively adding up one sign and reading them as the opposite desired sign. This can be done as the elements are being entered and they are all processed the same way for each sign, passed as it is or add it to a sum of other numbers of its sign before passing it, which is a linear time complex process. \textbf{O$(n)$}.
                \item Then the set has to be sorted by absolute value. Sorting algorithms such as Merge Sort can be implemented to efficiently sort the elements within worst case \textbf{O$(n\log n)$} time complexity.
                \item Adding these time complexities to the previously discussed auxiliary time complexities, we get \textbf{O$(C)$}, where '$C$' is sub-exponential worst case time. Average case \textbf{O$(C)$}, where '$C$' is polynomial and greater than linear. Best case \textbf{O$(n\log n)$}, where '$C$' is linear or less and sorting complexity dominates over complexities linear and under.
            \end{itemize}
        \item \textbf{Space Complexity}:
            \begin{itemize}
                \item[] \textbf{Auxiliary Space Complexity (only)}:
                \item The LIFO stack manages storage dynamically at runtime, but in the worst case, the amount of storage needed would not exceed linear space complexity (Section 3.4 and 3.7). \textbf{O$(n)$}.
                \item When zooming-in into a S/E point, we have to store the S/E point we zoomed into so that if we maybe do not find a solution in the inner S/E points, we are able to move back to higher level S/E points we zoomed into (LIFO stack). Since every level of S/E point enumeration uses constant space (section 3.7), and on each zoom-in we should remember where we left off in the higher level S/E points, each zoom-in introduces a constant amount of storage.
                \item The amount of zoom-ins that would need to be perform in the worst case of any arbitrary set grows half-linearly to the input size, \textbf{O$\left(\frac{n}{2}\right)$} (section 3.4). Therefore the amount of storage used in the worst case is the constant memory for each level, \textbf{O$(1)$}, multiplied by the amount of zoom-ins that need to be done in the worst case, which is half-linear, \textbf{O$\left(\frac{n}{2}\right)$}, giving us a linear space complexity.
                \item[] \[ \textbf{O}(1) \times \textbf{O}\left(\frac{n}{2}\right) = \textbf{O}\left(\frac{n}{2}\right) = \textbf{O}(n) \]
                \item Overall the auxiliary space complexity of the method is linear. \textbf{O$(n)$}.
                
                \item[] \textbf{Including Input Storage Space Complexity}:
                \item Firstly, for the overhead, it needs only a constant amount of storage, the count of zero elements. \textbf{O$(1)$} (section 3.9).
                \item The overhead also has to store the cumulative sum of the one sign if it is a mixed-sign set. This will help the method later when it is enumerating S/E points of the set. It is one variable and therefore, constant space complex. \textbf{O$(1)$}.
                \item Storing the elements of the set and the target sum takes linear storage \textbf{O$(n)$}.
                \item For sorting, an additional \textbf{O$(n)$} space may be used for sorting algorithms such as Merge Sort.
                \item Adding these complexities to the auxiliary space complexity, we get a linear space complexity. \textbf{O$(n)$}.
            \end{itemize}
        \item \textbf{Comparison With Traditional Methods}:
            \begin{itemize}
                \item \textbf{Method Presented}:
                \item[] Best-case and Average-case Time Complexity = \textbf{O$(C)$}, where '$C$' is the number of candidates and is a polynomial greater than linear and \textbf{O$(n \log n)$}.
                \item[] Worst-case Time Complexity = \textbf{O$(C)$}, for sub-exponential '$C$'.
                \item[] Space Complexity = \textbf{O$(n)$}
                
                \item \textbf{Dynamic Programming}:
                \item[] Best-case and Average-case Time Complexity = \textbf{O$(n \times s)$}, where '$n$' is the number of elements and '$s$' is the target sum. For best and average case, '$s$' and '$n$' should be of a small to moderate size.
                \item[] Worst-case Time Complexity = \textbf{O$(n \times s)$}, where '$n$' and '$s$' are of high magnitude and makes the performance of the algorithm degrade as more subsets need to be evaluated using the Dynamic Programming (DP) table.
                \item[] Space Complexity = \textbf{O$(n \times s)$}
                
                \item \textbf{Backtracking}:
                \item[] Best-case and Average-case Time Complexity = \textbf{O$\left(2^n\right)$}
                \item[] Worst-case Time Complexity = \textbf{O$\left(2^n\right)$}
                \item[] Space Complexity = \textbf{O$(n)$}
                
                \item \textbf{Meet in the Middle}:
                \item[] Best-case and Average-case Time Complexity = \textbf{O$\left(2^{\frac{n}{2}}\right)$}
                \item[] Worst-case Time Complexity = \textbf{O$\left(2^{\frac{n}{2}}\right)$}
                \item[] Space Complexity = \textbf{O$\left(2^{\frac{n}{2}}\right)$}
                
                \item \textbf{Branch and Bound}:
                \item[] Best-case and Average-case Time Complexity $<$ \textbf{O$(2^n)$}, depends on the pruning efficiency but less than exponential.
                \item[] Worst-case Time Complexity = \textbf{O$\left(2^n\right)$}
                \item[] Space Complexity = \textbf{O$(n)$}
            \end{itemize}
    \end{enumerate}
    
    Experimental results of the algorithm on different cases, including edge cases, of the Subset Sum Problem is demonstrated in section 4.
    

    \subsection{The Counting Complexity Barrier}
    \begin{theorem}
        Verifying false positives of any instance is equivalent to counting solutions if those false positives where solutions.

        \begin{enumerate}
            \item By definition of false positives, these are path traversed by the algorithm up to the very end where there should be solutions. However, when we get to the end of these paths, it is 2 adjacent possibilities that satisfy the sum-bound criteria but there is no further zooming in. So, the adjacent possibilities follow each other in the binary increment enumeration in a way that skips over the target sum very closely.
            \item In the smallest S/E point with distance 1 that satisfies the sum-bound criteria, it should be one of the points touching the target sum or none (Theorem 4).
            \item As this causes full exploration of a path, it would take exactly that exploration to reach a solution if it was possible at the end of that path.
            \item Therefore, for any number of false positives, validating them all would be the same as counting that many solutions if they were possible. As much as the algorithm cannot effciently count all solutions if there is only an exponential number of them, we also cannot effciently decide unsolvability in an instance with only an exponential number of false positives. For deciding other instances it takes a counting complex computation.
        \end{enumerate}
    \end{theorem}

    
    \subsection{Algorithm Implementation}
    To copy and reproduce algorithm implementation, please download the '.tex' file of this paper and copy the code from there.

    \begin{lstlisting}
        /********************************************************************
        **Author/Program by**: Thami Nkosi
        **Email Address**: 222133860@student.uj.ac.za
        **Program abstract**: This is a Subset Sum Problem Solver program. It illustrates in code (C++) the method presented in a research paper titled "Algorithmic Structure In Subset Sum: Deterministic In-Bound Navigation And The Counting Complexity Divide" by Thami Nkosi. It is a fresh approach to solving the Subset Sum Problem and consequently other NP/NP-complete problems. It overcomes most shortcomings of traditional methods, is optimizable and heuristic-friendly. Presenting, a structurally adaptive Deterministic Turing Machine for deciding and counting solutions to an NP-complete problem:

        The program goes through the whole possibility space looking for all solutions and counting them all along with elements of the set contributing to the solutions it finds. If there is no solution, the program will not display any solutions (may run into an exponential number of false positives in the worst case) and will indicate in the end that there are zero possible solutions. To counter exponential false positive validation, we let the method list all solutions even if there is an exponential number of them. If the purpose is just to decide solvability (finding at least one) and the program is taking time going through many solutions, simply press "Ctrl + C", and you will already have a number of solutions to your Subset Sum instance. If the program is able to run through the whole possibility space before you terminate it by pressing "Ctrl + C", you will have all the possible solutions (or none) to your Subset Sum instance and the count of the solutions found in the end. Catch it if you can.

        For clarity, refer back to sections in the paper referenced in the code. Program execution begins in the 'main' function.
        ********************************************************************/

        #include <iostream>
        #include <cmath>
        #include <algorithm>
        #include <vector>
        #include <cstdlib>

        using namespace std;

        //A "NavigateLevel" function prototype. Full declaration to be done later in the code.
        void NavigateLevel(vector<long long>&, unsigned long long, long long, signed long long, long long&, unsigned long long&);

        unsigned long long numOddElements; // A variable to help us recognize the trivial Subset Sum case where you are given an odd target sum and only even number elements. There is no solution in this instance.
        //********************************************************************/
        //A function that processes the elements in the set before they are passed for searching the target sum.
        //Sections to read: 3.4, 3.5 and 3.13.
        signed long long Preprocessor(vector<long long>& Elements, long long negElementsTotal) {

            //Searches for negative elements in the set and adds them all up. This sum is to be passed back to the caller of this function.
            //We always process the elements of a set as positive in this program, but it would not be incorrect to process it as negative or dynamically between negative and positive as desired. Simply adapt this code accordingly.
            for (long long& element : Elements) {
                if (element < 0)
                    negElementsTotal += element; //Linear time complex process (O(n)).
                if ((abs(element) % 2) == 1)
                    numOddElements++;
            }

            //Sort the elements in the set by their absolute value without modifying the sign of negative elements in the set.
            sort(Elements.begin(), Elements.end(), [](long long a, long long b) {
                return abs(a) < abs(b); //Linearithmic/Log-Linear time complex (O(nlogn)).
            });

            //Sorting is has the dominant complexity and therefore, the overall complexity of this function is Linearithmic/Log-Linear (O(nlogn)).
            return negElementsTotal;
        }
        //********************************************************************/

        //********************************************************************/
        //A function to map back to the elements in the set the positions of where the target sums are found.
        //The positions of the solutions are passed as parameter "numPreceding" into this function.
        //Section to read: 3.6.
        string elementCombinations(vector<long long>& Elements, unsigned long long numPreceding) {

            string contributingElements = ""; //Variable used to store the elements contributing to the solution in position "numPreceding".
            unsigned long long numElements = Elements.size() - 1;

            if (numPreceding > 0) {
                //Loop through the elements in the set and determine if the element contributes to sum to the solution in position "numPreceding".
                //When the element in the set is negative, the bit ("rem") corresponding to the element is negated.
                //This is a linear time complex process (O(n)).
                for (unsigned long long n = 0; n <= numElements; n++) {
                    if (numPreceding == 1 && Elements[n] > 0) {
                        //During the conversion of the position "numPreceding", when the position is "1", it should be taken as is and not divided further.
                        //If the element in the set corresponding to it is positive, we add the element to "contributingElements". Otherwise, we would have to negate it and would not add it to "contributingElements".
                        contributingElements == "" ? contributingElements += to_string(Elements[n]) : contributingElements += " + " + to_string(Elements[n]);
                        numPreceding = 0; //When "numPreceding" is equal to "1", it would mean the conversion has reached its highest bit to initially be set to "1" and all higher bits would trivially be set to "0".
                    } else if (numPreceding == 0 && Elements[n] < 0) {
                        //Negates the higher bits that were trivially set to "0" if the element corresponding to it is negative. The bit then turns to "1" and means the corresponding element should be added to "contributingElements".
                        contributingElements == "" ? contributingElements += to_string(Elements[n]) : contributingElements += " + " + to_string(Elements[n]);
                    } else {
                        int rem = numPreceding % 2;

                        //Negating bits that correspond to negative elements in the set.
                        //Helps us correctly identify if the element contributes to sum to the solution even for set with negative elements or a mix of negative and positive elements.
                        if (Elements[n] < 0 && rem == 1)
                            rem = 0;
                        else if (Elements[n] < 0 && rem == 0)
                            rem = 1;

                        if (rem == 1 && (n == 0 || contributingElements == ""))
                            contributingElements += to_string(Elements[n]);
                        else if (rem == 1)
                            contributingElements += " + " + to_string(Elements[n]);
                        
                        numPreceding /= 2;
                    }
                }
                
                if (contributingElements == "")
                    contributingElements += "Empty Set Zero";

            } else
                contributingElements += "Empty Set Zero";

            //The time complexity of this function is linear (O(n)).
            return contributingElements;
        }
        //********************************************************************/

        unsigned long long depth = 0, ActualCountSolutions = 0, PrecedingPossibilities = 0, TrivialCountSolutions = 0;
        //********************************************************************/
        //A function to determine whether the desired solution can or cannot be found within a S/E point.
        //Sections to read: 3.4, 3.6, 3.10 and 3.13.
        bool ZoomInCriteria(vector<long long>& Elements, long long& Target, long long num1, long long num2, unsigned long long num3, bool isFirstSEP, unsigned long long& AllSEPSC) {

            //"isFirstSEP" means "is First S/E point".
            unsigned long long currentPreceding = PrecedingPossibilities; //Stores the number of preceding possibilities for events where we zoom-in into a S/E point.

            if (num1 <= Target && num2 >= Target) {
                for (unsigned long long d = 0; d <= depth; d++) {cout << "\t";}
                cout << PrecedingPossibilities << "(" << num1 << ", " << num2 << ")" << exp2(num3 + 1) - 1 << endl;
                
                if ((num1 == Target || num2 == Target)) {
                    //If the starting or ending point of a S/E point are equal to the target sum, there is no need to zoom into it as all inner possibilities would be less than (if ending point = target) or greater than (if starting point = target) the target sum.
                    //The program outputs this as a solution and does not further zoom-in (with proper indentation of levels where the solution is found).
                    //This can also be a point where the program realizes the latest when it has run into a false positive if the execution flow does not reach this scope.
                    for (unsigned long long d = 0; d <= depth; d++) {cout << "\t";}
                    if (num1 == Target)
                        cout << "Target Sum found! = " << PrecedingPossibilities << "(" << num1 << ")0 = " << elementCombinations(Elements, PrecedingPossibilities) << endl;
                    else if (num2 == Target)
                        cout << "Target Sum found! = " << PrecedingPossibilities + exp2(num3 + 1) - 1 << "(" << num2 << ")0 = " << elementCombinations(Elements, PrecedingPossibilities + exp2(num3 + 1) - 1) << endl;
                    
                    ActualCountSolutions++, AllSEPSC++;
                } else {
                    if (num3 > 0) {
                        //We zoom-in deeper into a S/E point that satisfies the criteria and properly increase and decrease "depth".
                        //To zoom-in, we again call the "NavigateLevel" function, passing in specific values.
                        //NOTE! The number of elements ("num3"/second parameter of "NavigateLevel") we pass here are less than number of elements in the "NavigateLevel" that called this "ZoomInCriteria" function.
                        //Therefore, as "NavigateLevel" is linear time complex, passing a lesser number of elements to work with, makes the complexity of the "NavigateLevel" called here to be sub-linear.
                        //We pass in "0" as the "negElementsTotal" because there would be no need to again shift the inner possibilities.
                        depth++;
                        NavigateLevel(Elements, num3, num1, 0, Target, AllSEPSC);
                        depth--;
                        PrecedingPossibilities = currentPreceding; //Recalls the number of preceding possibilities before we zoomed in.
                    } else {
                        for (unsigned long long d = 0; d <= depth; d++) {cout << "\t";}
                        cout << "False Positive!" << endl; //Program realizes the latest that it has ran into a false positive otherwise, if it zooms back or skips exploration without reaching the deepest depth, it detected the false positive early and avoided zooming.
                        return true;
                    }
                }
                
                //In the pair of S/E points found using the current element in "NavigateLevel", if the S/E point processed in this "ZoomInCriteria" is the first S/E point of the pair and its starting point is equal to the target, there is no need to further explore any other possibilities to come after in this level.
                //Possibilities to come after will all be greater than the target sum.
                if (num1 > Target && isFirstSEP == true) {return false;}
                else {return true;}
            } else {
                //In the pair of S/E points found using the current element in "NavigateLevel", if the S/E point processed in this "ZoomInCriteria" is the first S/E point of the pair and its starting point is greater than the target, there is no need to further explore any other possibilities to come after in this level.
                //Possibilities to come after will all be greater than the target sum.
                if (num1 > Target && isFirstSEP == true) {return false;}
                else {return true;}
            }

            //Overall Complexity of this function is constant time complex (O(1)). The number of steps taken to verify if we should zoom-in or not, or continue exploring S/E points of this level or not are the same for any S/E point.
        }
        //********************************************************************/

        //********************************************************************/
        //A function to help us perform  the trivial count of redundant subsets as a product of repeating elements in the set.
        //Section to read: 3.10.
        void Trivial(vector<unsigned long long>& TrivialCountArr, unsigned long long TrivialArrSize, unsigned long long& TrivialTotal, bool skipTrivial) {

            bool AltAddSubtract = false; //Variable to alternate the addition and substraction of trivial terms.
            unsigned long long Coeffient = TrivialArrSize - 1;
            bool skipTrivialCount = skipTrivial;
            //Check if any of the subsets that were actually explored had solutions. If the have no solutions, there no trivial solutions to be calculated.
            //This check is only done for the highest level repeating element. For lower levels during which the function is cumulatively adding the "TrivialTotal", we take "skipTrivialCount" from the highest level repeating element.
            if (TrivialArrSize == TrivialCountArr.size()) {
                for (int i = 0; i < (TrivialArrSize - 1); i++) {
                    if (TrivialCountArr[i] != 0) {
                        skipTrivialCount = false;
                        break; //breakout of the loop if we find even one actual explored subsets count greater than 0.
                    }
                }
            }
            //Accordingly adds and subtracts the terms of each repeating element level all up to the highest level.
            if (skipTrivialCount == true) {
                for (unsigned long long d = 0; d <= depth; d++) {cout << "\t";}
                cout << "End Of Repetition! No Trivial Solutions Within, Trivial Count Up To Date." << endl;
            } else {
                for (int i = 1; i <= TrivialArrSize - 1; i++) {
                    if (TrivialArrSize == 2) {
                        AltAddSubtract = true;
                    }
                    if (Coeffient == 1) {
                        if (AltAddSubtract == false) {
                            TrivialTotal -= TrivialCountArr[0];
                            TrivialCountArr[i] -= TrivialCountArr[0];
                            AltAddSubtract = true;
                        } else {
                            TrivialTotal += TrivialCountArr[0];
                            TrivialCountArr[i] += TrivialCountArr[0];
                            AltAddSubtract = false;
                        }
                    } else if (i == 1) {
                        Trivial(TrivialCountArr, TrivialArrSize - 1, TrivialTotal, skipTrivialCount);
                        TrivialTotal += (Coeffient * TrivialCountArr[(TrivialArrSize - (i + 1))]);
                        TrivialCountArr[(TrivialArrSize - 1)] += (Coeffient * TrivialCountArr[(TrivialArrSize - (i + 1))]);
                        Coeffient *= ((TrivialArrSize - (i + 1)));
                        Coeffient /= (i + 1);
                    } else {
                        if (AltAddSubtract == false) {
                            TrivialTotal -= (Coeffient * TrivialCountArr[(TrivialArrSize - (i + 1))]);
                            TrivialCountArr[(TrivialArrSize - 1)] -= (Coeffient * TrivialCountArr[(TrivialArrSize - (i + 1))]);
                            AltAddSubtract = true;
                        } else {
                            TrivialTotal += (Coeffient * TrivialCountArr[(TrivialArrSize - (i + 1))]);
                            TrivialCountArr[(TrivialArrSize - 1)] += (Coeffient * TrivialCountArr[(TrivialArrSize - (i + 1))]);
                            AltAddSubtract = false;
                        }
                        Coeffient *= ((TrivialArrSize - (i + 1)));
                        Coeffient /= (i + 1);
                    }
                }
                if (TrivialArrSize == TrivialCountArr.size()) {
                    for (unsigned long long d = 0; d <= depth; d++) {cout << "\t";}
                    cout << "End Of Repetition! Trivial Count Updated." << endl;
                }
            }

            //Complexity of this function is O(n^2) if it had to calculate the trivial solutions within. Otherwise, it would be skipping over the n^2 possibilities.
        }
        //********************************************************************/

        //********************************************************************/
        //A function that successively moves through S/E points of each level.
        //Sections to read: 3.3, 3.5, 3.7, 3.10 and 3.13.
        void NavigateLevel(vector<long long>& Elements, unsigned long long numElements, long long StartingPoint, signed long long negElementsTotal, long long& Target, unsigned long long& AllSEPSC) {

            //"AllSEPSC" means All S/E point Solutions Count. It is incremented for all trivial and actual solutions count.
            unsigned long long FirstSEPSC = 0, SecondSEPSC = 0; //Suffix "SEPSC" means "S/E Point Solutions Count".
            long long num1 = 0, num2 = 0, num3 = 0, num4 = 0, num5 = 0;
            unsigned long long num6 = 0;
            unsigned long long lastPartSEPdistance = 0, lastPartSEPSC = 0; //Variables to store the distance of subsets we would have to explore in case of repeating elements and the number of solutions found within those subsets.
            long long RepetitionCount = 0;
            bool Continue = true; //A variable to check if we should continue exploring S/E points of this level.
            bool isRepeating = false, ReportRepetitionDetectionOnce = false; //"ReportRepetitionDetectionOnce" is a variable to make the program output the "Repetition Detected" message only once from the start of repetition and its end on each level.
            vector<unsigned long long> TrivialCountArr;
            //From each element in the set we get a pair of S/E points, except for the first 2 elements in the set. So, :
            /*
                num1 = starting point of the first S/E point.
                num2 = ending point of the first S/E point.
                num3 = starting point of the second S/E point.
                num4 = ending point of the second S/E point.
                num5 = utility variable to help us move across S/E points.
                num6 = tells us the distance/number of possibilities represented by a S/E point. It only stores the number of elements enough to explore the S/E point so, calculations can be performed using it to achieve other needs in the program.
            */
            //Looping through the elements of the set bounded by the parameter "numElements" is linear time complex and sub-linear if "numElements" is less than the number of elements in the set.
            for (unsigned long long element = 0; element <= numElements && Continue == true; element++) {
                //Shifts and translations to variable num1 to num4 are applied accordingly using the sum of "StartingPoint" and "negElementsTotal".
                //As the negative elements in the set were not changed sign but ordered by their absolute value, when accessing them, we read them in their absolute values as we need the S/E points we get to be of a single-sign set that we later shift.
                if (element == 0) {
                    //S/E point using the first element in the set.
                    num1 += StartingPoint + negElementsTotal;
                    num2 = abs(Elements[element]) + StartingPoint + negElementsTotal;
                    Continue = ZoomInCriteria(Elements, Target, num1, num2, num6, true, FirstSEPSC);
                    AllSEPSC += FirstSEPSC;
                } else if (element == 1) {
                    //S/E point using the second element in the set.
                    num3 = abs(Elements[element]) + StartingPoint + negElementsTotal;
                    num4 = abs(Elements[element]) + abs(Elements[element - 1]) + StartingPoint + negElementsTotal;
                    PrecedingPossibilities += 2;
                    Continue = ZoomInCriteria(Elements, Target, num3, num4, num6, false, SecondSEPSC);
                    AllSEPSC += SecondSEPSC;
                } else if (element == 2) {
                    FirstSEPSC = 0, SecondSEPSC = 0;
                    //Variables num1 to num4 are translated accordingly.
                    //The third element is used and it is the first element that gives us a pair of S/E points.
                    num1 = abs(Elements[element]) + StartingPoint + negElementsTotal;
                    num2 = abs(Elements[element]) + abs(Elements[element - 2]) + StartingPoint + negElementsTotal;
                    num3 = abs(Elements[element]) + abs(Elements[element - 1]) + StartingPoint + negElementsTotal;
                    num4 = abs(Elements[element]) + abs(Elements[element - 1]) + abs(Elements[element - 2]) + StartingPoint + negElementsTotal;
                } else {
                    //We successively move across S/E points performing arithmetic operations only on variables num1 to num5.
                    //Here, each element always produces a pair of S/E points.
                    num5 = abs(Elements[element]) + StartingPoint + negElementsTotal;

                    if (num5 == num1) 
                        isRepeating = true;

                    num2 = (num4 - num1) + num5;
                    num3 = num1 + num5 - (StartingPoint + negElementsTotal);
                    num4 += num5 - (StartingPoint + negElementsTotal);
                    num1 = num5;
                    num6++;

                    if (isRepeating == true) {
                        //Increments the "RepetitionCount" accordingly as it detects repetition in the next S/E point pair and begins passing the count of solutions we found in the previous S/E point pair into "TrivialCountArr".
                        RepetitionCount == 0 ? RepetitionCount = 2 : RepetitionCount++;
                        if (TrivialCountArr.size() == 0) {
                            TrivialCountArr.push_back(FirstSEPSC + SecondSEPSC);
                            FirstSEPSC = 0, SecondSEPSC = 0;
                            lastPartSEPdistance = num6 + 1;
                        }
                    } else {
                        //Detects end of repetition then calls "Trivial" to perform the trivial count of solutions in subsets skipped over.
                        if (TrivialCountArr.size() != 0){
                            unsigned long long intermediateTrivialTotal = 0;
                            Trivial(TrivialCountArr, TrivialCountArr.size(), intermediateTrivialTotal, true);
                            TrivialCountArr = vector<unsigned long long>();
                            TrivialCountSolutions += intermediateTrivialTotal;
                            AllSEPSC += intermediateTrivialTotal;
                            ReportRepetitionDetectionOnce = false;
                        } else 
                            FirstSEPSC = 0, SecondSEPSC = 0;
                        
                        lastPartSEPdistance = 0, RepetitionCount = 0;
                    }
                }

                //Prevents checking of S/E points got from the first 2 elements, and delegated that task to the if statement above.
                if (element > 1 && isRepeating == false) {
                    //Passes the first S/E point in the pair to the 'ZoomInCriteria' function.
                    //Increments Preceding Possibilities accordingly using "num6".
                    num6 == 0 ? PrecedingPossibilities += 2 : PrecedingPossibilities += exp2(num6);
                    Continue = ZoomInCriteria(Elements, Target, num1, num2, num6, true, FirstSEPSC);
                    
                    //If "ZoomInCriteria" has flagged that the program should not continue exploring S/E point of this level in the first S/E point, the check of the second S/E point can be skipped.
                    if (Continue == true) {
                        num6 == 0 ? PrecedingPossibilities += 2 : PrecedingPossibilities += exp2(num6 + 1);
                        Continue = ZoomInCriteria(Elements, Target, num3, num4, num6, false, SecondSEPSC);
                    }
                    AllSEPSC += (FirstSEPSC + SecondSEPSC);
                } else if (isRepeating == true) {
                    //Calls the "ZoomInCriteria" function passing in the appropriate paramters to explore the non-redundant subsets in case of repeating elements.
                    //Passes the count of solutions found in the non-redundant subsets exploration into "TrivialCountArr"
                    if (ReportRepetitionDetectionOnce == false) {
                        for (unsigned long long d = 0; d <= depth; d++) {cout << "\t";}
                        cout << "Repetition Detected! Trivial Count To Be Updated When Repetition Ends." << endl;
                        ReportRepetitionDetectionOnce = true;
                    }
                    PrecedingPossibilities += (exp2(num6) + (2 * exp2(num6 + 1)) - exp2(lastPartSEPdistance));
                    Continue = ZoomInCriteria(Elements, Target, (abs(Elements[num6 + 1]) * RepetitionCount) + (StartingPoint + negElementsTotal), num4, lastPartSEPdistance - 1, false, lastPartSEPSC);
                    PrecedingPossibilities -= (exp2(num6 + 1) - exp2(lastPartSEPdistance));
                    TrivialCountArr.push_back(lastPartSEPSC);
                    AllSEPSC += lastPartSEPSC;
                    lastPartSEPSC = 0;
                }
                isRepeating = false;
            }
            //Detects the end of repetition at the end of a level
            if (TrivialCountArr.size() != 0) {
                unsigned long long intermediateTrivialTotal = 0;
                Trivial(TrivialCountArr, TrivialCountArr.size(), intermediateTrivialTotal, true);
                TrivialCountArr = vector<unsigned long long>();
                TrivialCountSolutions += intermediateTrivialTotal;
                AllSEPSC += intermediateTrivialTotal;
                ReportRepetitionDetectionOnce = false;
            }

            //The overall complexity of this function is independantly linear time complex for each level in the LIFO stack.
        }
        //********************************************************************/

        //********************************************************************/
        //Program execution begins here in the "main" function.
        // Sections to read: 3.3, 3.4, 3.5, 3.6, 3.7, 3.8, 3.9, 3.10 and 3.13.
        int main() {

            vector<long long> Elements;
            long long enteredElement, NumZeros = 0, TargetSum;
            unsigned long long numElements = 0;
            signed long long negElementsTotal = 0;
            bool isPureZeroInstance = false; //Pure Zero Instance is when the Subset Sum instance has a target sum equal to zero and containing of only zeros in its set. This variable helps the program keep count of the solutions equal to zero for a pure zero Subset Sum instance.

            cout << "What is the target Sum?: ";
            cin >> TargetSum;

            while (true) {
                for (long long element : Elements)
                    cout << element << "; ";
                
                if (Elements.size() == 0)
                    cout << "Enter element: ";
                else
                    cout << endl << "Enter element: ";

                cin >> enteredElement;

                if (enteredElement == 0)
                    NumZeros++;
                else {
                    Elements.push_back(enteredElement);
                    numElements++;
                }

                char choice;
                cout << endl << "End of Input?(y/n): ";
                cin >> choice;
                if (choice == 'y' || choice == 'Y') {
                    #ifdef _WIN32
                        system("cls");
                    #else
                        system("clear");
                    #endif
                    break;
                } else {
                    #ifdef _WIN32
                        system("cls");
                    #else
                        system("clear");
                    #endif
                }
            }

            negElementsTotal = Preprocessor(Elements, negElementsTotal);
            cout << "Elements: ";
            for (long long element : Elements)
                cout << element << "; ";

            cout << endl << "Target Sum: " << TargetSum << endl;

            unsigned long long TotalNumSolutions = 0; //This is a variable that is incremented for every solution that is found, i.e. both trivial and actual count. This variable is/should be equal to "TrivialCountSolutions" plus "ActualCountSolutions" at the end of the program.
            if (TargetSum == 0 && Elements.size() == 0) {
                cout << "Pure Zero Subset Sum Instance! Subset Sum Instance Has Target Sum Zero And A Set Containing Only Zeros." << endl;
                isPureZeroInstance = true;
            } else if (numOddElements == 0 && (TargetSum % 2) == 1)
                cout << "All Even Number Elements Cannot Result To An Odd Target Sum." << endl;
            else
                NavigateLevel(Elements, numElements - 1, 0, negElementsTotal, TargetSum, TotalNumSolutions);

            //Number of solutions found has to be multiplied with the number of zero elements entered into the program as zeros cause "z-axis exponentiation" and increases number of solutions.
            //"isPureZeroInstance" variable is used in an addition expression in the following line of code even though it is a boolean data type. If "true", it is added in the expression as "1" and if "false", as "0". It helps the program be able to keep count of the number of solutions even in a pure zero Subset Sum instance.
            cout << endl << "Zeros Count: " << NumZeros << endl << "Trivially Counted Solutions: " << TrivialCountSolutions << endl << "Actual Count Of Solutions Before Processing: " << ActualCountSolutions << endl << "Number Of Possible Solutions Found: " << (TotalNumSolutions + isPureZeroInstance) * exp2(NumZeros) << " = ((" << ActualCountSolutions << " + " << TrivialCountSolutions << ") * 2^(" << NumZeros << "))";

            cout << endl << "End of Program" << endl;
            return 0;
        }
        //********************************************************************/
    \end{lstlisting}


\section{Experimental Results}
These experimental results demonstrate the efficiency of this method in solving the decision version of different edge cases of the subset sum problem. We do not consider the cumulative preceding possibilities here because we just want to determine if the solution is possible or not.

\begin{enumerate}
    \item \textbf{Large numbers with small differences}. Sets containing large integers that are close in value can make it difficult to identify which subset might sum to a target.
    \item[] Given: \{1000000, 1000001, 1000002, 1000003, 1000004\} and Target Sum: 3000006 
    \item[] \textbf{Solution}:
        \begin{itemize}
            \item The set is already ordered, so we determine the S/E points: 0, 1000000, 1000001, 2000001, (1000002, 2000002)\textsubscript{1}, (2000003, 3000003)\textsubscript{1}, (1000003, 3000004)\textsubscript{3}, (2000005, 4000006)\textsubscript{3}, (1000004, 4000007)\textsubscript{7}, (2000007, 5000010)\textsubscript{7}.
            \item All positive set therefore, criteria is: $S <= 3000006$ and $E >= 3000006$
            \item First S/E point that satisfies the criteria: (2000005, 4000006)\textsubscript{3}.
            \item Translate the first pattern that fits into this S/E point: (2000005, 4000006)\textsubscript{3} = 2000005, 3000005, 3000006, 4000006.
            \item We have found a possibility that is equal to 3000006.
            \item Therefore, the solution of 3000006 for the given set exists.
        \end{itemize}
    \item \textbf{Sets with Repeated Elements}: Sets with repeated elements, especially when the repeating elements are more in number compared to unique elements.
    \item[] Given: \{5, 5, 5, 5, 10\} and Target Sum: 20
    \item[] \textbf{Solution}:
        \begin{itemize}
            \item The set is ordered, single-sign and has no zeros.
            \item Determine the S/E points: 0, 5, 5, 10, (5, 10)\textsubscript{1}, (10, 15)\textsubscript{1}, (5, 15)\textsubscript{3}, (10, 20)\textsubscript{3}, (10, 25)\textsubscript{7}, (15, 30)\textsubscript{7}.
            \item A possibility of 20 is already the ending point of S/E point: (10, 20)\textsubscript{3}
            \item Therefore, the solution of 20 in the given set exists.
        \end{itemize}
    \item \textbf{Sparse and Dense Combinations}: Sets containing both very small and very large numbers.
    \item[] Given: \{1, 6, 31, 16, 8, 4, 2\} and Target Sum: 30
    \item[] \textbf{Solution}:
        \begin{itemize}
            \item Order the set: \{1, 2, 4, 6, 8, 16, 31\}.
            \item Determine S/E points: 0, 1, 2, 3, (4, 5)\textsubscript{1}, (6, 7)\textsubscript{1}, (6, 9)\textsubscript{3}, (10, 13)\textsubscript{3}, (8, 15)\textsubscript{7}, (14, 21)\textsubscript{7}, (16, 29)\textsubscript{15}, (24, 37)\textsubscript{15}, (31, 52)\textsubscript{31}, (47, 68)\textsubscript{31}.
            \item All positive set therefore, the criteria is: $S <= 30$ and $E >= 30$
            \item First S/E point that satisfies the criteria: (24, 37)\textsubscript{15}
            \item Translate first pattern that fits into the S/E point: (24, 31)\textsubscript{15} = 24, 25, 26, 27, (28, 29)\textsubscript{1}, (30, 31)\textsubscript{1}, (30, 33)\textsubscript{3}, (34, 37)\textsubscript{3}.
            \item We have found a possibility that is equal 30.
            \item Therefore, a solution of 30 for the given set exists.
        \end{itemize}
    \item \textbf{Negative and Positive Mixed Sets}: Sets containing a mix of both negative and positive numbers, which cancel each other out.
    \item[] Given: \{-1, 2, -3, 4, -5, 6\} and Target Sum: 3
    \item[] \textbf{Solution}:
        \begin{itemize}
            \item Add all the negative elements and get their sum: $-1 + (-3) + (-5) = -9$
            \item Order the set by absolute value and read them all as positive: \{-1, 2, -3, 4, -5, 6\}.
            \item Determine S/E points:  0, 1, 2, 3, (3, 4)\textsubscript{1}, (5, 6)\textsubscript{1}, (4, 7)\textsubscript{3}, (7, 10)\textsubscript{3}, (5, 11)\textsubscript{7}, (9, 15)\textsubscript{7}, (6, 16)\textsubscript{15}, (11, 21)\textsubscript{15}
            \item Shift by “-9”:  -9, -8, -7, -6, (-6, -5)\textsubscript{1}, (-4, -3)\textsubscript{1}, (-5, -2)\textsubscript{3}, (-2, 1)\textsubscript{3}, (-4, 2)\textsubscript{7}, (0, 6)\textsubscript{7}, (-3, 7)\textsubscript{15}, (2, 12)\textsubscript{15}
            \item We read it as all positive therefore, criteria is: $S <= 3$ and $E >= 3$
            \item First S/E point that satisfies the criteria: (0, 6)\textsubscript{7}
            \item Translate first pattern that fits into the S/E point: (0, 6)\textsubscript{7} = 0, 1, 2, 3, (3, 4)\textsubscript{1}, (5, 6)\textsubscript{1}.
            \item We have found a possibility that is equal to 3.
            \item Therefore, a solution of 3 for the set exists.
        \end{itemize}
    \item \textbf{Close to Half-Sum Target}: Sets where the target sum is very close to half the sum of all elements. Most subsets tend to sum to numbers close to half the sum of all the elements, which makes it harder to find a subset that has a target sum close to the half-sum.
    \item[] Given: \{2, 3, 5, 7, 11\} and Target Sum: 14
    \item[] \textbf{Solution}:
        \begin{itemize}
            \item Set is ordered, single-sign and has no zeros.
            \item Determine S/E points: 0, 2, 3, 5, (5, 7)\textsubscript{1}, (8, 10)\textsubscript{1}, (7, 12)\textsubscript{3}, (12, 17)\textsubscript{3}, (11, 21)\textsubscript{7}, (18, 28)\textsubscript{7}.
            \item All positive set therefore, the criteria is: $S <= 14$ and $E >= 14$
            \item First S/E point that satisfies the criteria: (12, 17)\textsubscript{3}
            \item Translate first pattern that fits into the S/E point: (12, 17)\textsubscript{3} = 12, 14, 15, 17.
            \item We have found a possibility that is equal to 14.
            \item Therefore, a solution of 14 for the set exists.
        \end{itemize}
    \item \textbf{Nearly All Elements required}: Sets where the target sum is close to the sum of all but few elements of the set.
    \item[] Given: \{1, 2, 4, 8, 16, 31\} and Target Sum: 59
    \item[] \textbf{Solution}:
        \begin{itemize}
            \item Set is ordered, single-sign and has no zeros.
            \item Determine S/E points: 0, 1, 2, 3, (4, 5)\textsubscript{1}, (6, 7)\textsubscript{1}, (8, 11)\textsubscript{3}, (12, 15)\textsubscript{3}, (16, 23)\textsubscript{7}, (24, 31)\textsubscript{7}, (31, 46)\textsubscript{15}, (47, 62)\textsubscript{15}.
            \item All positive set therefore, the criteria is: $S <= 59$ and $E >= 59$
            \item First S/E point that satisfies the criteria: (47, 62)\textsubscript{15}
            \item Translate first pattern that fits into the S/E point: (47, 62)\textsubscript{15} = 47, 48, 49, 50, (51, 52)\textsubscript{1}, (53, 54)\textsubscript{1}, (55, 58)\textsubscript{3}, (59, 62)\textsubscript{3}.
            \item We have found a possibility that is equal to 59.
            \item Therefore, a solution of 59 for the given set exists.
        \end{itemize}
    \item \textbf{Power of Two Sets}: Sets containing powers of 2. These sets widen the normal distribution of the set increasing possibility of subsets achieving more sums.
    \item[] Given: \{1, 2, 4, 8, 16\} and Target Sum: 14
    \item[] \textbf{Solution}:
        \begin{itemize}
            \item Set is ordered, single-sign and has no zeros.
            \item Determine the S/E points:  0, 1, 2, 3, (4, 5)\textsubscript{1}, (6, 7)\textsubscript{1}, (8, 11)\textsubscript{3}, (12, 15)\textsubscript{3}, (16, 23)\textsubscript{7}, (24, 31)\textsubscript{7}.
            \item All positive set therefore, the criteria is: $S <= 14$ and $E >= 14$
            \item First S/E point that satisfies the criteria: (12, 15)\textsubscript{3}
            \item Translate first pattern that fits into the S/E point: (12, 15)\textsubscript{3} = 12, 13, 14, 15.
            \item We have found a possibility that is equal 14.
            \item Therefore, a solution of 14 for the given set exists
        \end{itemize}
    \item \textbf{High-Density Target Range}: Sets with many small numbers and a target sum that is slightly larger than the sum of large subsets of the set. These small numbers make the search space large.
    \item[] Given: \{1, 1, 1, 1, 1, 1, 1, 1, 1, 10\} and Target Sum: 9
    \item[] \textbf{Solution}:
        \begin{itemize}
            \item Set is ordered, single-sign and has no zeros.
            \item Determine the S/E points: 0, 1, 1, 2, (1, 2)\textsubscript{1}, (2, 3)\textsubscript{1}, (1, 3)\textsubscript{3}, (2, 4)\textsubscript{3}, (1, 4)\textsubscript{7}, (2, 5)\textsubscript{7}, (1, 5)\textsubscript{15}, (2, 6)\textsubscript{15}, (1, 6)\textsubscript{31}, (2, 7)\textsubscript{31}, (1, 7)\textsubscript{63}, (2, 8)\textsubscript{63}, (1, 8)\textsubscript{127}, (2, 9)\textsubscript{127}, (10, 18)\textsubscript{255}, (11, 19)\textsubscript{255}.
            \item We have found a possibility of 9 in the S/E point (2, 9)\textsubscript{127}. The ending point is equal to 9.
            \item Therefore, a solution of 9 for the given set exists.
        \end{itemize}
    \item \textbf{Random Large Sets}: A large set of random integers, especially when the target sum is set randomly close to but not equal any of the large elements in the set.
    \item[] Given: \{17, 25, 31, 47, 59, 71, 89, 97\} and Target Sum: 166
    \item[] \textbf{Solution}:
        \begin{itemize}
            \item Set is ordered, single-sign and has no zeros.
            \item Determine the S/E points: 0, 17, 25, 42, (31, 48)\textsubscript{1}, (56, 73)\textsubscript{1}, (47, 89)\textsubscript{3}, (78, 120)\textsubscript{3}, (59, 132)\textsubscript{7}, (106, 179)\textsubscript{7}, (71, 191)\textsubscript{15}, (130, 250)\textsubscript{15}, (89, 268)\textsubscript{31}, (160, 339)\textsubscript{31}, (97, 347)\textsubscript{63}, (186, 436)\textsubscript{63}.
            \item All positive set therefore, the criteria is: $S <= 166$ and $E >= 166$
            \item First S/E point that satisfies the criteria: (106, 179)\textsubscript{7}
            \item Translate first pattern that fits into the S/E point: (106, 179)\textsubscript{7} = 106, 123, 131, 148, (137, 154)\textsubscript{1}, (162, 179)\textsubscript{1}.
            \item S/E point that satisfies the criteria is (162, 179)\textsubscript{1}, but the distance between this S/E point is 1, the starting and ending point are next to each other and there is no number in between. Section 3.15 program would output this as a false positive. We move to the next base S/E points.
            \item Next S/E points of the set that satisfies the criteria: (71, 191)\textsubscript{15}
            \item Translate first patterns that fit into this S/E point: (71, 191)\textsubscript{15} =  71, 88, 96, 113, (102, 119)\textsubscript{1}, (127, 144)\textsubscript{1}, (118, 160)\textsubscript{3}, (149, 191)\textsubscript{3}
            \item S/E point that satisfies the criteria: (149, 191)\textsubscript{3}
            \item Further translating first pattern that fit into this S/E point: (149, 191)\textsubscript{3} =  149, 166, 174, 191
            \item We have found a possibility of 166.
            \item Therefore, a solution of 166 for the given set exists.
        \end{itemize}
    \item \textbf{Miscellaneous Set}: 
    \item[] Given: \{1, 2, 4, 8, 16, 100, 100, 100, -36, 0\} and Target Sum: 190
    \item[] \textbf{Solution}:
        \begin{itemize}
            \item Add all negative elements in the set: -36.
            \item Remove 0 from the set.
            \item Order the set by absolute value and read them all as positive: \{1, 2, 4, 8, 16, -36, 100, 100, 100\}.
            \item Determine S/E points: 0, 1, 2, 3, (4, 5)\textsubscript{1}, (6, 7)\textsubscript{1}, (8, 11)\textsubscript{3}, (12, 15)\textsubscript{3}, (16, 23)\textsubscript{7}, (24, 31)\textsubscript{7}, (36, 51)\textsubscript{15}, (52, 67)\textsubscript{15}, (100, 131)\textsubscript{31}, (136, 167)\textsubscript{31}, (100, 167)\textsubscript{63}, (200, 267)\textsubscript{63}, (100, 267)\textsubscript{127}, (200, 367)\textsubscript{127}.
            \item Shift by “-36”: -36, -35, -34, -33, (-32, -31)\textsubscript{1}, (-30, -29)\textsubscript{1}, (-28, -25)\textsubscript{3}, (-24, -21)\textsubscript{3}, (-20, -13)\textsubscript{7}, (-12, -5)\textsubscript{7}, (0, 15)\textsubscript{15}, (16, 31)\textsubscript{15}, (64, 95)\textsubscript{31}, (100, 131)\textsubscript{31}, (64, 131)\textsubscript{63}, (164, 231)\textsubscript{63}, (64, 231)\textsubscript{127}, (164, 331)\textsubscript{127}.
            \item We read them as all positive therefore, the criteria is: $S <= 190$ and $E >= 190$
            \item First S/E point that satisfies the criteria: (164, 231)\textsubscript{63}
            \item Translate first pattern that fits into this S/E point: (164, 231)\textsubscript{63} = 164, 165, 166, 167, (168, 169)\textsubscript{1}, (170, 171)\textsubscript{1}, (172, 175)\textsubscript{3}, (176, 179)\textsubscript{3}, (180, 187)\textsubscript{7}, (188, 195)\textsubscript{7}, (200, 215)\textsubscript{15}, (216, 231)\textsubscript{15}
            \item First inner S/E point to satisfy the criteria is: (188, 195)\textsubscript{7}
            \item Translate first pattern that fits into this S/E point: (188, 195)\textsubscript{7} =  188, 189, 190, 191, (192, 193)\textsubscript{1}, (194, 195)\textsubscript{1}
            \item We have found a possibility equal to 190.
            \item Therefore, there is a solution of 190 in the given set.
        \end{itemize}
    \item \textbf{Miscellaneous Set}:
    \item[] Given: \{-15, -5, 0, 3, 6, 9, 11, 14, 20, 25, 25, 33, 40, 45, 60\} and Target Sum: 68
    \item[] \textbf{Solution}:
        \begin{itemize}
            \item Remove zero.
            \item This time, let us intentionally make the set all negative. Sum up all positive signed elements: $3 + 6 + 9 + 11 + 14 + 20 + 25 + 25 + 33 + 40 + 45 + 60 = 291$.
            \item Order the set by absolute value and read them all as negative: \{3, -5, 6, 9, 11, 14, -15, 20, 25, 25, 33, 40, 45, 60\}.
            \item We read them as all negative therefore, the criteria is: $S >= 68$ and $E <= 68$
            \item Determine the S/E points, because: 0, -3, -5, -8, (-6, -9)\textsubscript{1}, (-11, -14)\textsubscript{1}, (-9, -17)\textsubscript{3}, (-15, -23)\textsubscript{3}, (-11, -24)\textsubscript{7}, (-20, -34)\textsubscript{7}, (-14, -37)\textsubscript{15}, (-25, -48)\textsubscript{15}, (-15, -49)\textsubscript{31}, (-29, -63)\textsubscript{31}, (-20, -68)\textsubscript{63}, (-35, -83)\textsubscript{63}, (-25, -88)\textsubscript{127}, (-45, -108)\textsubscript{127}, (-25, -108)\textsubscript{255}, (-50, -133)\textsubscript{255}, (-33, -141)\textsubscript{511}, (-58, -166)\textsubscript{511}, (-40, -173)\textsubscript{1023}, (-73, -206)\textsubscript{1023}, (-45, -211)\textsubscript{2047}, (-85, -251)\textsubscript{2047}, (-60, -266)\textsubscript{4095}, (-105, -311)\textsubscript{4095}.
            \item Shift by “291”: 291, 288, 286, 283, (285, 282)\textsubscript{1}, (280, 277)\textsubscript{1}, (282, 274)\textsubscript{3}, (276, 268)\textsubscript{3}, (280, 267)\textsubscript{7}, (271, 257)\textsubscript{7}, (277, 254)\textsubscript{15}, (266, 243)\textsubscript{15}, (276, 242)\textsubscript{31}, (262, 228)\textsubscript{31}, (271, 223)\textsubscript{63}, (256, 208)\textsubscript{63}, (266, 203)\textsubscript{127}, (246, 183)\textsubscript{127}, (266, 183)\textsubscript{255}, (241, 158)\textsubscript{255}, (258, 150)\textsubscript{511}, (233, 125)\textsubscript{511}, (251, 118)\textsubscript{1023}, (218, 85)\textsubscript{1023}, (246, 80)\textsubscript{2047}, (206, 40)\textsubscript{2047}, (231, 25)\textsubscript{4095}, (186, 20)\textsubscript{4095}.
            \item First S/E point that satisfies the criteria: (206, 40)\textsubscript{2047}.
            \item Translate first patterns that fit into this S/E point: (206, 40)\textsubscript{2047} =  206, 203, 201, 198, (200, 197)\textsubscript{1}, (195, 192)\textsubscript{1}, (197, 189)\textsubscript{3}, (191, 183)\textsubscript{3}, (195, 182)\textsubscript{7}, (186, 172)\textsubscript{7}, (192, 169)\textsubscript{15}, (181, 158)\textsubscript{15}, (191, 157)\textsubscript{31}, (177, 143)\textsubscript{31}, (186, 138)\textsubscript{63}, (171, 123)\textsubscript{63}, (181, 118)\textsubscript{127}, (161, 98)\textsubscript{127}, (181, 98)\textsubscript{255}, (156, 73)\textsubscript{255}, (173, 65)\textsubscript{511}, (148, 40)\textsubscript{511}.
            \item First inner S/E point that satisfies the criteria is: (173, 65)\textsubscript{511}
            \item Translate first pattern that fits into this S/E point: (173, 65)511 =  173, 170, 168, 165, (167, 164)\textsubscript{1}, (162, 159)\textsubscript{1}, (164, 156)\textsubscript{3}, (158, 150)\textsubscript{3}, (162, 149)\textsubscript{7}, (153, 139)\textsubscript{7}, (159, 136)\textsubscript{15}, (148, 125)\textsubscript{15}, (158, 124)\textsubscript{31}, (144, 110)\textsubscript{31}, (153, 105)\textsubscript{63}, (138, 90)\textsubscript{63}, (148, 85)\textsubscript{127}, (128, 65)\textsubscript{127}.
            \item Next inner S/E point that satisfies the criteria is: (128, 65)\textsubscript{127}.
            \item Translate first pattern that fits into this S/E point: (128, 65)\textsubscript{127} =  128, 125, 123, 120, (122, 119)\textsubscript{1}, (117, 114)\textsubscript{1}, (119, 111)\textsubscript{3}, (113, 105)\textsubscript{3}, (117, 104)\textsubscript{7}, (108, 94)\textsubscript{7}, (114, 91)\textsubscript{15}, (103, 80)\textsubscript{15}, (113, 79)\textsubscript{31}, (99, 65)\textsubscript{31}.
            \item Next inner S/E point that satisfies the criteria is: (99, 65)\textsubscript{31}.
            \item Translate first pattern that fits into this S/E point: (99, 65)\textsubscript{31} =  99, 96, 94, 91, (93, 90)\textsubscript{1}, (88, 85)\textsubscript{1}, (90, 82)\textsubscript{3}, (84, 76)\textsubscript{3}, (88, 75)\textsubscript{7}, (79, 65)\textsubscript{7}.
            \item Next inner S/E point that satisfies the criteria is: (79, 65)\textsubscript{7}.
            \item Translate first pattern that fits into this S/E point: (79, 65)\textsubscript{7} =  79, 76, 74, 71, (73, 70)\textsubscript{1}, (68, 65)\textsubscript{1}
            \item We have found a possibility of 68.
            \item Therefore, a solution of 68 for the given set exists.
        \end{itemize}
\end{enumerate}


\section{Implications And Discussions}
Now, the capabilities of the method not covered extensively in the paper. Exploration of each S/E point level is independent of other levels. The only common thing is that they are explored by translations of the same set elements fitting the level. This makes the algorithm naturally parallelizable. A queue of processors/threads can be assigned to solve an instance, queue size scalable, and each S/E point level/branch can be explored by one processor. During a level's exploration, a processor would have two objectives; invoke the next processor in the queue to explore deeper any S/E points that satisfy the sum-bound criteria within the level and can be zoomed into, or reach a candidate itself if no further zoom-in can be done on the S/E point. After both those outcomes and/or reaching the end of a level, the processor goes back to the queue and await next call. Space complexity should also be considered in practical implementation.

The algorithm is also capable of simultaneously searching for multiple target sums. On every S/E point, the criteria would simply be a sum-bound check for all the target sums searched for instead of one. However, this would also increase the number of candidates. We can also be able to attribute the kind of solution we want, like looking for a solutions that would have specific elements. This can simply be done by skipping to where the highest desired element is introduced in the possibility space, and then only move by the position multiples of when it first appears. This reduces the number of candidates.

We did not refer much to Observation 2 in the paper, but it can also serve as a heuristic that could help find solutions faster for other instances. Other heuristics and/or approximations can also be implemented to reduce the number of candidates explored, further sharping the guiding mechanism. We also have intentions of improving the complexity of first candidate reach and candidate to candidate traversal to linear time worst case in future work.

Development of reductions according to the framework detailed in this paper are also worth further exploring. The framework shows promise in improving reduction complexities as we saw with the reduction from k-SAT and general CNF-SAT. We have committed ourselves to extensively detailing the generalization of this algorithm to k-SAT and general CNF-SAT in a follow-up paper.

This algorithm is structurally aware and adaptive as its complexities are governed only by the number of candidates. Inefficiency of the algorithm reflects the exact structural complexity of the instance itself. Other structurally blind algorithms are therefore bound to run into exponential behaviour as even a structurally aware algorithm runs into exponential behaviour in the worst case. An algorithm would have to be structurally aware and further prune through the in-bound solution space efficiently in order to be able to efficiently decide worst case instances. That would be the only way the counting complexity barrier would collapse by a way that explicity explores subsets. Either than that, a mathematical concept would have to be discovered to collapse the counting complexity barrier. Only by those two ways would we be able to resolve the P vs NP in either way. So, the presented algorithm can quite efficiently decide and count solutions, and even in the worst case, efficiently keep stepping towards completely resolving the problem instance eventually even if exponential. With sufficient parallelization, the algorithm practically shrinks the intractability gap.

We are also dedicated to further study the exact upper bounds of the number of candidates and characterizing its worst case instances. This calls for a deeper understanding of combinatorics between diverse elements. An equivalent quantum algorithm is another research avenue worth exploring.


\section{Conclusion}
Though the algorithm handles most Subset Sum instance with impressive practical efficiency, its performance amplified by parallel computing, it remains theoretically exponential. It lacks a counting complementary part, a missing piece attainable through algorithmic optimization or mathematical insight. Whether such a construct exists or not would not only improve the algorithm, but resolve the P vs NP problem in either direction. Until then, intractability continues to practically shrink, while theoretical hardness endures.




\begin{thebibliography}{99}

    \bibitem{bellman1952}
    R.~Bellman.
    \newblock On the theory of dynamic programming.
    \newblock {\em Proceedings of the National Academy of Sciences}, 38(8):716--719, 1952.
    
    \bibitem{cook1971}
    S.~Cook.
    \newblock The complexity of theorem-proving precedures.
    \newblock In {\em Proceedings of the Third Annual ACM Aymposium on Theory of Computing (STOC '71)}, pages 151--158. ACM, 1971.
    
    \bibitem{horowitz1974}
    E.~Horowitz and S.~Sahni.
    \newblock Computing partitions with applications to the knapsack problem.
    \newblock {\em Journal of the ACM}, 21(2):277--292, 1974.
    
    \bibitem{johnson1974}
    D.~S. Johnson.
    \newblock Approximation algorithms for combinatorial problems.
    \newblock {\em Journal of Computer and System Sciences}, 9(3):256--278, 1974.
    
\end{thebibliography}
\end{document}